\documentclass[12pt,preprint]{aastex}
\usepackage{natbib}
\shortauthors{Bower et al.}
\shorttitle{Radio Transients}
\begin{document}

\def\gsim{\;\rlap{\lower 2.5pt
 \hbox{$\sim$}}\raise 1.5pt\hbox{$>$}\;}
\def\lsim{\;\rlap{\lower 2.5pt
   \hbox{$\sim$}}\raise 1.5pt\hbox{$<$}\;}

\newcommand\degd{\ifmmode^{\circ}\!\!\!.\,\else$^{\circ}\!\!\!.\,$\fi}
\newcommand{\etal}{{\it et al.\ }}
\newcommand{\uv}{(u,v)}
\newcommand{\rdm}{{\rm\ rad\ m^{-2}}}
\newcommand{\msuny}{{\rm\ M_{\sun}\ y^{-1}}}
\newcommand{\mylesssim}{\stackrel{\scriptstyle <}{\scriptstyle \sim}}
\newcommand{\sci}{Science}


\title{SubmilliJansky Transients in Archival Radio Observations}

\author{Geoffrey C. Bower, Destry Saul, Joshua S. Bloom\altaffilmark{1}, 
Alberto Bolatto, Alexei V. Filippenko, Ryan J. Foley, and Daniel Perley \\
Department of Astronomy \& Radio Astronomy Laboratory,
University of California, Berkeley, CA 94720-3411; gbower@astro.berkeley.edu}
\altaffiltext{1}{Sloan Research Fellow.}

\begin{abstract}
We report the results of a 944-epoch survey for 
transient sources with archival data from the Very Large
Array spanning 22 years with a typical epoch separation
of 7 days.  Observations were obtained at 5 or 8.4~GHz
for a single field of view with a full-width at half-maximum 
of $8.6^\prime$ and $5.1^\prime$, respectively,
and achieved a typical point-source detection threshold at the beam center of
$\lsim 300 \ \mu$Jy per epoch. The angular resolution ranged from $\sim 1''$ 
to $15''$. Ten transient sources were detected with a significance
threshold such that only one false positive would be expected.  
Of these transients, eight
were detected in only a single epoch.  Two transients were
too faint to be detected in individual epochs but were detected in
two-month averages.  None of the 
ten transients was detected in longer-term averages or associated 
with persistent emission in the deep image produced from the combination 
of all epochs.  The cumulative
rate for the short timescale radio transients above 370 $\mu$Jy at 5
and 8.4 GHz is
$0.07 \lsim R \lsim 40 {\rm\ deg^{-2} yr^{-1}}$, where the uncertainty
is due to the unknown duration of the transients,
$20 {\rm\ min} \lsim t_{char} \lsim 7 {\rm\ d}$.  A two-epoch 
survey
for transients will detect $1.5 \pm 0.4$ transient per square degrees above
a flux density of 370 $\mu$Jy.
One transient is located $\sim 3$ kpc in projection from the nucleus of 
a spiral galaxy at a redshift $z = 0.040$.  Based on the
duration of the transient, its luminosity ($L \approx 2 \times 10^{38}$
erg s$^{-1}$), and its location in the galaxy, we suggest
that it may be similar to the peculiar Type Ib/c radio supernova 
SN 1998bw associated with GRB 980428.  The implied Type Ib/c rate is 
consistent to an order of magnitude
with rates determined optically.  A second transient is associated
with a blue galaxy at $z = 0.249$ and may also be a luminous radio
supernova or gamma-ray burst afterglow.  Two other transient
sources are possibly associated with the outer parts of 
faint, optically detected galaxies.  The remaining six transients
have no counterparts in the optical or infrared to limiting magnitudes
of $R \approx 27$ and $K_s \approx 18$ and show no faint persistent
radio flux.  For the eight transients without clear associations, 
known source classes of radio transients
including radio supernovae, gamma-ray burst afterglows, 
active galactic nuclei, and stellar flares do not provide natural 
fits to the observed parameters. The hosts and progenitors of these 
transients are unknown.

\end{abstract}

\section{Introduction}

Radio transient sources have primarily been studied 
through follow-up observations after discovery at optical, 
X-ray, or $\gamma$-ray wavelengths
\citep[e.g.,][]{1971Natur.234..138H,1973A&A....26..105D,
1997Natur.389..261F,2002ApJ...573..306E,1999Natur.398..127F,
2005Natur.434.1104G}, or through serendipitous discovery 
\citep[e.g.,][]{1976Natur.261..476D,1992Sci...255.1538Z,2003ApJ...598.1140B}.
This is principally due to the high cost of observing
time to survey large areas of sky to significant depth at radio
wavelengths \citep{2004NewAR..48.1459C}.  
Moreover, existing blind surveys have typically been performed at
low frequencies, where
survey time is shorter but the effect of synchrotron self-absorption
may hide certain source classes.  
The available parameter space for radio transient surveys is extensive:
transients have been detected at, and are predicted for all, radio 
wavelengths; timescales of transient and variable behavior range
from nanoseconds \citep{2003Natur.422..141H} to the longest timescales 
probed \citep{1992ApJ...396..469H}; and transients may originate from 
nearly all astrophysical environments including the solar system 
\citep{2003PASP..115..675K}, star-forming regions
\citep{2003ApJ...598.1140B}, the Galactic
Center \citep{1992Sci...255.1538Z,2005ApJ...633..218B}, and other
galaxies \citep{1997Natur.389..261F}.

Nevertheless, extensive surveys have been carried out, exploring different
volumes of the parameter space, and, in some cases, discovering new phenomena.
Roughly, one can separate transients surveys into two classes: (a) burst
searches that probe timescales of less than $\sim 1$ second, are
often performed with low angular resolution instruments, and are most often
performed at low frequencies; and (b) imaging surveys
conducted with interferometers which typically probe timescales of tens of
seconds and longer.  Examples of burst searches include
STARE, a 611-MHz all-sky survey \citep{2003PASP..115..675K} sensitive
to timescales of 0.1~s to a few minutes; a 843-MHz survey with
the Molonglo Observatory synthesis Telescope (MOST) 
sensitive to timescales of 1 $\mu$s
to 800 ms \citep{1989PASAu...8..172A}; 
and the Parkes 1.4-GHz multi-beam pulsar survey.
STARE used dipole antennas with a sensitivity threshold
of 27 kJy simultaneously over a significant fraction of the sky
and found no extra-solar transients on timescales of 0.125~s
to a few minutes in 18 months of observations.  The MOST survey
determined an upper limit to the transient rate of 5 events 
${\rm s^{-1}\ deg^{-2}}$ for 10~ms events at a flux density
limit of 1~Jy.  The Parkes multi-beam survey has been used for extensive
searches for periodic emission as well as 
for pulsed emission.  The latter search led to the discovery of
rotating radio transients (RRATs), which appear to be pulsars with
erratic emission \citep{2006Nature...mcl}.  

Imaging surveys for transients have focused on individual fields
and on large-scale surveys.
Recently, a blind survey of the Galactic center at 330 MHz
has discovered an unusual radio transient that displays 
quasi-periodic emission on a timescale of 70 min
\citep{2005Natur.434...50H}.  \citet{2003ApJ...590..192C} 
used $\sim 30$~hr of observation of the Lockman hole to characterize 
variability on a timescale of 19~d and 17 mon at 1.4 GHz.  They found
no transients above 100 $\mu$Jy and conclude that the transient density
on these timescales is fewer than 18 per square degree.  
Similarly,
\citet{2003AJ....125.2299F} identified 4 highly variable sources but no
transient sources at 5 and
8.4 GHz using VLA observations of fields in which gamma-ray bursts are
present.  They set an upper limit to the variable source population of 6
deg$^{-2}$ at sub-mJy sensitivity, comparable to the limit set by
\citet{2003ApJ...590..192C}.  Recently, drift scan interferometric
observations have identified several transients of unknown origin, 
including a 3-Jy transient with a duration of 72
hr and three $>1$ Jy transients  of $i\lsim 1$~d duration
\citep{2007ApJ...657L..37N,2007PASP..119..122K,2007AJ....133.1441M}.

The most extensive imaging search for transients derives from a comparison
of the FIRST and NVSS 1.4~GHz catalogs
\citep{2002ApJ...576..923L,2006ApJ...639..331G}. The search covered
2400 square degrees at a flux density threshold of 6 mJy.  This search
generated a number of transient candidates. Of these, one 
source was identified as a radio supernova in a nearby galaxy and another
was seen only once in the radio and could not be 
associated with any known optical source.  Transient identification
in this survey suffers from the 
methodological problem of the extremely mismatched resolutions (FIRST:  
$5^{\prime\prime}$, NVSS:  $45^{\prime\prime}$)  of the 
two surveys.

In addition to radio supernovae (RSNe), a number of highly variable and
transient sources are expected to be detectable in significant
numbers at radio wavelengths.  One example is 
gamma-ray burst (GRB) afterglows associated with GRBs which are beamed
in a direction away from Earth, and therefore not observable at
high-energy wavelengths \citep{1997ApJ...487L...1R}.  
These so-called ``orphan gamma-ray burst afterglows'' (OGRBAs)
are expected to outnumber GRBs by a factor of 10 to 1000,
the specific value depending on the characteristics of the relativistic
outflow of the GRB and the surrounding medium.  There may be related 
phenomena such as those associated with X-ray flashes
\citep{2005ApJ...629..311S}, supernovae (SNe)
\citep{2003ApJ...599..408B}, and short-duration GRBs 
\citep{2005Natur.437..851G}.
Activity from stars and compact objects in the
Galaxy may also be detected as transients 
\citep[e.g.,][]{2001Natur.410..338B,2005ApJ...633..218B,Osten06}.

A range of active galactic nucleus (AGN) phenomena is also
anticipated, including intra-day variability and other short-term 
variability driven by the effect of interstellar scintillation 
on compact components of extragalactic radio sources 
\citep[e.g.,][]{2003AJ....126.1699L}.
Intrinsic AGN variability is also expected.  Sudden increases in 
the mass accretion rate following the tidal disruption of an orbiting 
star, for instance, will produce a transient soft X-ray flare and
possibly radio emission \citep{1988Natur.333..523R,2006IAUS..238E..97G}.

Significant numbers of RSNe, OGRBAs, and tidal flares 
are predicted for surveys that achieve a sensitivity $\sim 0.1$~mJy and 
cover an area of $\sim 10$ square degrees.  These phenomena all
produce self-absorbed synchrotron radiation and, therefore, are more
likely to be observed at high radio frequencies.

We describe here the results of a search for extragalactic radio transients
using 5 and 8.4 GHz observations from the Very Large Array data archives.  
The survey includes 944 epochs over 22 years consisting of observations
of the same field.  At a sensitivity threshold of $370$ $\mu$Jy,
the survey has an effective survey area comparable to that of a 
survey consisting of two epochs of 10 square degrees each,
making this thus far the largest-area sub-mJy imaging survey of 
radio transient phenomena.

\section{Observations and Analysis}

\subsection{The Archival Survey}

Between 1983 and 2005, the VLA performed calibration and system check 
observations 
of the same blank field with a typical interval of $\sim 7$~d.  
Each epoch typically was 20 min in duration and consisted of 10~s 
integrations. The field is centered at 
$\alpha =15^{h}02^{m}20.53^{s}$,  $\delta = +78^\circ16'14.905''$ 
(J2000).  The field is
out of the plane of the Galaxy ($l = 115^\circ$, $b = 36^\circ$).
626 epochs of observations were obtained at a frequency of 
5 GHz between 1983 and 1999.
599 epochs were obtained at a frequency of 8.4 GHz between 1989 and 2005.
VLA observations of this field ceased in 2005.
Between 1989 and 1999, there was overlap between some of the 5 and 8.4 GHz
observations.  281 of the 8.4 GHz epochs were simultaneously observed
at 5 GHz, giving a total of 944 independent epochs at either frequency.

Data were all obtained in standard continuum mode with 50 MHz of bandwidth
in each of two intermediate-frequency bands (with centers separated by
100 MHz) and two circular polarizations.
We used a pipeline procedure for flagging, calibrating, and imaging.
Phase calibration was made based on 
brief observations of the compact source J1803+784.  Since no
standard flux density calibrator was used in these observations, we 
set the amplitude scale of the observations by assuming the mean
flux density for J1803+784
as measured in the University of Michigan Radio Astronomy Observatory 
database over the same period \citep[2.2 Jy and 2.8 Jy 
at 5 and 8.4 GHz, respectively][]{1999ApJ...512..601A}.
The assumption of constant flux density introduces 
less than 15\% uncertainty in
the flux-density scale for individual epochs.  Images of the target 
field typically had a root-mean-square (rms) error in the flux density 
of 40 to 50 $\mu$Jy.
Images from two successive epochs (19840613 and 19840620) 
are shown in Figure~\ref{fig:typicalimage}.  Throughout the paper, we 
use the notation of YYYYMMDD to denote epoch date and refer to specific
transients as RT YYYYMMDD, based on the epoch of detection.

Images were obtained in all configurations of the Very Large Array.   
For the extended configurations, bandwidth smearing within the primary
beam can become significant.  We compensated for this effect by
applying a Gaussian taper with a full-width at half-maximum (FWHM) 
of 150 $k\lambda$
to the $\uv$ data before imaging, which reduces sensitivity and
angular resolution but
provides imaging to large radii without bandwidth distortions.  
Images were made using natural
weighting to achieve maximal sensitivity and were deconvolved using
the CLEAN algorithm \citep{1980A&A....89..377C}.  We imaged a region slightly larger
than the region containing
a circle of radius equal to twice the half-power radius, which is $9.0'$
at 5 GHz and $5.4'$ at 8.4 GHz.

A deep image of the region at 5 and 8.4 GHz was also constructed through 
merging of
all visibilities and imaging with a $\uv$ taper of 150 $k\lambda$
using the MIRIAD software package.  The resulting images had resolutions
of $\sim 5''$ and $\sim 3''$, and an rms of 2.6 $\mu$Jy 
and 2.8 $\mu$Jy at 5 and 8.4 GHz, respectively.   The location
of steady and transient radio sources on the deep image is shown
in Figure~\ref{fig:deep}.  Fluxes in the deep image are typically lower
than the average of detections since our high detection threshold and
the low flux densities of the sources bias
the mean.  Further details of the deep images will be
published separately.  

\subsubsection{Transient Identification}

Sources were identified using AIPS task SAD (search and destroy), which
identifies peaks in the image above a flux-density threshold and fits
point sources.  For the relatively uncluttered images that we have, SAD
is efficient and accurate.
We identified sources within twice the half-power radius and corrected
their flux densities for primary beam losses.

The detection threshold for each image was chosen such that the probability
of a false detection (PFD) in the full imaged region was $10^{-3}$, or
a single false source for 1000 identical epochs, assuming Gaussian statistics
for noise in the image.  
Since the synthesized beam changes with array configuration while
the primary beam remains the same, the number
of independent pixels in the field of view varied substantially.
Detection thresholds were in the range $\sim 5\sigma$ to $6\sigma$, 
dependent on array configuration.  The typical flux density threshold at
the center of the image was $\sim 300$ $\mu$Jy. For the number of epochs 
observed and the circular area of the region searched for sources, we have
an expectation of $\sim 1$ false source in the entire survey.
Inverting the sign of the intensity in each image 
and searching for sources with the same thresholds and methods revealed
one significant source, consistent with our expectations of one or fewer
false detections in the entire survey.

Wider-field images ($\sim 40'$ at 5 GHz, $\sim 27'$ at 8.4 GHz)
were made for all epochs in which a transient was detected.
A number of transients that were detected on the western and northern edges 
of the individual epoch image were determined to be aliased power from two
bright radio sources ($\sim 50$ mJy) at radii greater than two
primary beam widths (NVSS J1457+7817, J1500+7827).  
One source above the detection threshold was rejected on the basis of
a non-pointlike image and deep negative stripes in the image.

Detected sources and their measured properties are listed in 
Tables~\ref{tab:steady} and \ref{tab:transients}.
Table~\ref{tab:steady} identifies persistent sources above or near
the detection threshold and not multiple-epoch transients.
Six persistent sources at 5 GHz and 2 persistent sources at 8.4 GHz
were detected in multiple epochs.  The epochs of detection were
distributed over the entire survey.  
Of these persistent sources,
three radio sources with flux densities $> 500$ $\mu$Jy were
detected hundreds of times.  
One radio source (J150123+781806) was detected with
our pipeline software at 5 GHz in 
452 of 626 epochs, and at 8.4 GHz in 159 of 599 epochs
(Figure~\ref{fig:standardlightcurve}); with a reduced source detection
threshold of $3\sigma$, we find that this source was present in 602 epochs 
at 5 GHz and in 445 epochs at 8.4 GHz.  The results for this source indicate
the overall quality of the data.
Variations in the flux density of this source are caused by the 
changing sensitivity of the VLA
in different configurations, variability in the 
amplitude calibrator J1803+784,
and possibly intrinsic variability in the source.

Seven sources at 5 GHz and 
1 source at 8.4 GHz were detected in individual epochs
(Table~\ref{tab:transients}).
These sources were detected
only once and are classified as transients.
We plot the light curves of each transient for the year
surrounding their detection (Figure~\ref{fig:lct}). 

We searched for time variability for each transient by splitting the
data into four-minute segments and comparing the flux densities.  
The reduced $\chi^2$ for the hypothesis of no variability is less than 1.4 
for all the transients, implying no strong evidence for variability.  We
also imaged each transient in Stokes V and found no evidence for circular
polarization above $\sim 30\%$.  Finally, we differenced images
between the intermediate frequency bands and found no evidence for 
a large spectral index.

The radial distribution of sources is consistent with a real source population
rather than imaging defects.  Imaging defects 
would be uniformly distributed throughout the image, without regard 
for the decline in sensitivity due to the primary beam.  
Four of the eight transients are contained within the
half-power radius.  This number is 
inconsistent with the expectation of a uniform
distribution within the same region, $2.0 \pm 1.4$.
On the other hand,
a source population with a power-law distribution of flux densities 
$N(S) \propto
S^{-3/2}$ in a flux-density limited sample has an expectation of 
$3.6 \pm 1.3$ within the half-power radius, in better agreement with the 
observed number.

Images were also constructed in a search for faint radiation from the 
transients and for new transients using all data from two-month, 
one-year, and one-decade time spans.  We found two new transients in 
the two-month averages (Table~\ref{tab:transientsmonths}) 
and no transients in the year and decade images.
Light curves for these sources are shown in Figure~\ref{fig:lct}.  Given
the smaller number of epochs, the expected number of false sources in 
the two-month average data is $\sim 0.1$.  Further searches on different
time scales or with filters matched to light curve shapes
are possible with the same data but are beyond the scope of this paper.

Dual frequency observations were made only after 1989 and not at every
epoch.  Of the remaining 5 GHz transients, only 1 (RT 19920826)
is located within the imaged region of the 8.4 GHz survey.  
Unfortunately, there were no 8.4 GHz observations 
for this epoch.  Similarly, for RT 19970205 which
was detected at 8.4 GHz, there was not a simultaneous 5 GHz observation.

\subsection{Keck Imaging}

On 25 July 2006 (UT dates are used throughout this paper), 
we observed the central portion of the
radio field with the Low-Resolution Imaging Spectrometer
\citep[LRIS;][]{Oke95} on the Keck-I 10-m telescope. Eight of the
radio transient positions (see Figure 6) were observed in two separate
pointing centers (each covered by the LRIS field-of-view of $6' \times
7.8'$). At each pointing center we obtained six dithered
10-min exposures. The instrument is outfitted with a beam splitter,
allowing us to observe simultaneously in the $g$ and $R$ bands. The data
were reduced following standard optical imaging reduction
procedures. After fitting a world coordinate system (tied to the USNO
B1.0) to each reduced frame, we made a mosaic for a given filter using
{\sc SWARP} \citep{2002ASPC..281..228B}.  The final astrometric 
uncertainty relative to the International
Celestial Reference System (ICRS) is 250 mas in each axis. 

On the night of the Keck/LRIS imaging, we observed the  Landolt 
standard star field PG1633+099 \citep{1992AJ....104..340L} in  
both filters at an airmass of $\sec z = 1.04$. The catalog 
magnitudes  were converted to the $g$ band from the $V$-band
magnitude and the $B-V$ 
color using the prescription in Bilir et al.\ (2005) 
\nocite{2005AN....326..321B}.  From the observed photometry of 
four PG1633+099 stars, the zeropoint  conversion from flux in both 
filters was determined in an $0.6''$ aperture. The flux 
of  $3\sigma$-detected sources in both the $R$ and $g$ stacked 
image of the radio field were determined in the same-sized  
aperture. Accounting for the average extinction per unit airmass at 
Mauna Kea, we determine that the non-detections of some of the 
counterparts correspond to upper limits of 
$g({\rm lim}) \approx 27.6$ and 
$R ({\rm lim}) \approx 26.55$ mag. 
These limits assume an unresolved  source and a color 
similar to those of the Landolt field stars.

\subsection{Keck Spectroscopy}

An optical spectrum of the counterpart of RT 19840613 was taken 
on 2005 Dec. 4
with the DEIMOS spectrograph \citep{Faber03} mounted on the Keck-II
telescope, using a 600 line mm$^{-1}$ grating and the GG400 order-blocking
filter. Spectra of the
counterparts and nearby companions of 
RTs 19860115, 19870422, 19920826, and 19970528 were taken on
2006 June 26 with LRIS using the 600/4000 grism, the 400/8500
grating, and the D560 dichroic.

All spectral data were reduced using standard techniques 
\citep[e.g.,][]{2003PASP..115.1220F}.  Standard CCD
processing and spectrum extraction were completed with IRAF\footnote{IRAF
is distributed by the National Optical Astronomy Observatory, which is
operated by the Association of Universities for Research in Astronomy,
Inc., under cooperative agreement with the National Science Foundation.}.
The data were extracted with the optimal algorithm of \citet{Horne86}.
We obtained the wavelength scale from low-order polynomial fits to
calibration-lamp spectra.  Small wavelength shifts were applied after
cross-correlating night-sky lines extracted with the object to a template
sky.  Using our own IDL routines, we fit spectrophotometric standard star
spectra to flux calibrate our data and remove telluric lines \citep{Wade88,
Matheson00}.  

We give redshifts for the counterparts and nearby companions in 
Table~\ref{tab:redshift}.  In the cases of RTs 19840613, 19870422, 19920826,
and 19970528, the redshift is for the galaxy that the RT is situated in
or closest to.  RT 19860115 has two possible hosts in the Keck image;  
Objects 1 and 2 are located to the northeast and southwest of the RT,
respectively.

\subsection{PAIRITEL Observations}

Data were obtained with PAIRITEL \citep{2006ASPC..351..751B}
on 16 Feb. 2006. Images were taken
simultaneously in the $J$, $H$, and $K_s$ bands in four different pointings 
with a total integration time of about 1120~s per pointing.  
Additionally, to get a deep infrared (IR) image in one
portion of the radio field, several hours of imaging in June and July
2006 were obtained centered at J2000 position $\alpha$ = 
$15^{h}02^{m}36.0^{s}$, $\delta$ = $+78^{\circ}16'39''$. The 
approximate  $3\sigma$ upper limits
in the shallow, wide-angle map are $J = 19.2$ mag, $H = 18.5$ mag,
and $K_s = 18.0$ mag (the limits are significantly worse toward the
edges of the mosaic). 
The pointings were stitched together using SWARP after
finding preliminary WCS solutions in all bands. The result is
a $1050 \times 1050$ pixel set of mosaic images with $1'' \times 1''$
resolution that overlaps most of the radio image.  
An astrometric solution for
the mosaic images was found through a
cross-correlation with 43 stars in common with the USNO B1.0
catalog. Typical rms uncertainty of the astrometric tie to the ICRS
is 250 mas.

\section{Results}

We have detected a total of 8 transient radio sources in single epochs 
of this survey (Figure~\ref{fig:lct}).  Additionally, we have detected
two transients in two-month averages of the data.
Our source identification methods described previously
argue that no more than one of these sources is a false
detection.   All of these sources 
are adequately fit as point sources.  None of the transients is
detected in epochs other than the detection epoch, or in longer-term
averages with the possible exception of RT 19840613, in which the
host galaxy is detected in the full 20-year image.

 In Figures~\ref{fig:deepradio} and ~\ref{fig:keck} we show 
images of the transients overlaid on the deep radio image, on
multi-color images from the Keck imaging, and, where Keck data
are missing for two objects, on $K_s$-band images from PAIRITEL.

We identify several hosts or potential hosts to the RTs using 
the Minnesota Palomar Plate Survey 
\citep[MAPS, http://aps.umn.edu;][]{1993PASP..105..521P}.  MAPS
is a catalog of sources found with and characterized by
plate-measuring equipment surveying
the Palomar Sky Survey I (POSS I).  MAPS
coordinates are consistent with the radio reference frame with an
accuracy of $0.3''$.  We confirmed this by comparison of the position of
two bright reference stars in the MAPS catalog with positions in
the Tycho catalog.  Referring the Tycho positions to epoch 1950 (appropriate
for POSS), we find
agreement of $\sim 0.1''$ for GSC 04562-00668 and $\sim 1''$ for
GSC 04562-00458.

\subsection{A Transient Associated with a Spiral Galaxy: RT 19840613}

   One transient is of particular interest because it is coincident with
a spiral galaxy at a redshift $z = 0.040$ (a distance of $\sim 170$ Mpc
using $H_0 = 72 {\rm km ^{-1} Mpc^{-1}}$).  The galaxy, identified as 
MAPS-P023-0189928, has $R=16.5$ mag, $B-R=1.8$ mag, and a major axis 
of $\sim 20''$. The transient is clearly non-nuclear; it is offset to 
the southwest of the optical nucleus by $2.7'' \pm 0.4''$, corresponding 
to a projected distance of $2.3 \pm 0.3$ kpc.  The optical nucleus
is coincident with the infrared nucleus as identified in the 2MASS
catalog at an accuracy of $0.2''$.  The probability of a single chance
association with an optical source in the POSS at a few arcsecond 
precision is $\lsim 1\%$.  We calculate an isotropic radio luminosity of 
$2 \times 10^{28}$ erg s$^{-1}$ Hz$^{-1}$, or $2 \times 10^{38}$ erg s$^{-1}$
for a bandwidth of 10 GHz.

The transient is not detected in any other individual epoch, even with
reduced detection thresholds.  The transient is also not detected in any
two-month average in the year surrounding its detection, nor is it detected
in any annual average from the entire data set. It is also not detected 
in the year prior to detection, nor in the six years following detection.  
The absence of a detection in other light curves at 5 GHz 
is consistent with a rapid rise and a decline
in the flux density with a power law of $t^{-\alpha},\alpha > 0.5$.  

The deep image of the region surrounding the transient shows extended
radio continuum emission associated with the spiral galaxy 
and a companion galaxy
(Figure~\ref{fig:840613deep}).  The location of the transient coincides with
a region of peak emission, but it is not possible to conclude whether that
emission is the result of the transient or a peak in the galactic 
emission.  The peak of emission
near the position of the transient is $\sim 28~ \mu$Jy, which corresponds to 
a star-formation rate of $<0.1 M_\sun\ {\rm yr}^{-1}$
\citep[e.g.,][]{2002ApJ...568...88Y}.

The association of the transient with the non-nuclear region of a spiral
galaxy, together with its luminosity, suggest that the transient may 
be a RSN. The transient is very unlikely to be a Type II RSN, as these
sources exhibit rise and decay timescales of months to years, implying
that a Type II RSN would be detected multiple times in our survey
\citep{2002ApJ...573..306E}.
Type Ia RSN have never been detected at radio wavelengths
\citep{2004NewAR..48.1377W}.  Type Ib/c
RSN, on the other hand, evolve on timescales of days to months.  The
anomalous Type Ib/c RSN associated with SN~1998bw and GRB~980425
doubled its flux density during its rise and halved its flux density
following the peak in $<10$~d 
\citep{1998Natur.395..663K}.
The light curve declined as $t^{-1}$
with the exception of a brief period of increase attributed to
clumpiness in the circumstellar medium.  The luminosity of
RT 19840613 places it among the brightest SNe~Ib/c,
although a factor of a few fainter than SN~1998bw
\citep{2003ApJ...599..408B}.  
The luminosity of SN 2002ap, also a SN~Ib/c,
is four orders of magnitude less than that of SN 1998bw
\citep{2003ApJ...599..408B}.  
The luminosity of RT 19840613 is three orders
of magnitude less than the typical radio luminosity of a GRB
afterglow.

\subsection{A Transient Associated with a Blue Galaxy: RT 19870422}

RT 19870422 is located within $\sim 1.5''$ of the centroid of
MAPS-P023-0189613.  The Keck image shows that this 
RT is clearly associated 
with a blue galaxy and is significantly offset from the nucleus.  
The integrated $R$ magnitude of the galaxy is 20.2, 
with $B-R = 2.5$ mag, and the
redshift is determined to be $z = 0.249$ (or a distance of 1050 Mpc).  
Galaxy-template matching to the Keck spectrum
shows it to most likely be an Sc galaxy with 
strong [\ion{O}{2}] and [\ion{O}{3}] emission, indicating current
star formation.

The inferred isotropic radio luminosity is 
$L_\nu \approx 7 \times 10^{29} {\rm\ erg\ s^{-1}\ Hz^{-1}}$, or 
$L \approx 7 \times 10^{39} {\rm\ erg\ s^{-1}}$ for a bandwidth of
10 GHz.  
RT 19870422 was detected in a two-month integration, implying a 
total energy release
$E \approx 4 \times 10^{46}$ erg.  We see no evidence for detection
in other epochs, implying a source that fades with a timescale of 
two months or less following its peak.
In the deep integration, we see that the galaxy is detected
with an irregular morphology (Figure~\ref{fig:870422deep}).  
The faint radio emission, however, is offset from the position of
RT 19870422.

RT 19870422 shares many characteristics with RT 19840613.  
The RT position is clearly separated from the peak of the optical emission, 
ruling out the possibility of an AGN event.  Its longer duration
and higher luminosity, however, differentiate it.  
The luminosity is nearly an order of magnitude greater 
than that of SN 1998bw, and two orders of magnitude more luminous
than the bright Type II SN 1979C \citep{2002ApJ...573..306E}.  
Thus, the luminosity falls in between the maximum observed
luminosity of a RSN and the typical luminosity of a
GRB afterglow.  The two-month duration of RT 19870422 is more consistent 
with known RSNe of both Type Ib/c or Type II, as well as GRB afterglows.  

\subsection{Transients Possibly Associated with Galaxies or 
Galaxy Groups:  RT 19920826 \& 19970528 }

RT 19920826 is located $\sim 5''$ to the southeast of
a red galaxy identified in the Keck image, which is not identified
in the MAPS catalog.
The host is a red object with no clear emission or absorption
lines.

RT 19970528 is located $\sim 5''$ to the northwest of 
MAPS-P023-0189499, which has $R=19.6$ mag and $B-R=2.4$ mag.  
Two other galaxies are identified within a radius
of $30''$.  
The nearest one is a red, early-type galaxy at $z = 0.245$.
Galaxy-template matching shows it is most likely an elliptical galaxy,
but it is also well fit by an S0 spectrum.

We computed
the probability of false association of these galaxies with 
transients through Monte Carlo simulations.  We randomly distributed 100,000
radio sources over the field and then measured the distance to
the nearest galaxy as identified in the MAPS catalog.  The probability
of a false association at a distance of less than $5''$ is only 2\%.
The probability of a false association at a distance of $15''$ is
$\sim 20$\%.  The probability for any one of the 10 RTs being randomly 
associated with
a galaxy at a distance of $5''$ is $\sim 20\%$, and at a distance of
$15''$ the probability is nearly unity.  The probability of multiple sources
located within $5''$ is $\lsim 5\%$.  Thus, the association of 
either RT 19970528 or RT 19920826 or both RTs with a galaxy is 
probable but not certain.

\subsection{Sources without Radio or Optical Counterparts:  RT 19840502, 
19860115, 19860122, 19970205, 19990504, \& 20010331}

There is no evidence for faint radio, optical, or infrared counterparts 
for six of our sources: RT 19840502, 19860115, 19860122, 19970205, 
19990504, \and 20010331. There are no sources apparent in the image 
obtained by averaging the succeeding two months of data, as well as 
in the year-long and decade-long averages.  
These results require a rapidly decreasing light curve, proportional
to $t^{-1}$ or steeper.

RT 19860115 has two possible hosts. The first possible host, Object 1, 
is located $\sim 15''$ to the northeast of the RT and is a
starburst galaxy at $z = 0.1297$ with strong emission lines and a blue
continuum.  
Object 1 is identified as 
MAPS-P023-0190130, a galaxy with $R=19.2$ mag and $B-R=1.3$ mag,  and
is also detected as a faint, extended radio source.
The other possible host, Object 2, is located $\sim 20''$ to
the southwest of the RT and is an early-type galaxy at
$z = 0.242$.  Galaxy-template matching shows the spectrum is best fit by
either an elliptical or S0 galaxy spectrum.
Several other objects, likely to be galaxies,
are in the field.  
The physical separation of RT 19860115 from the two objects is 
$\sim 50$--100 kpc, which is still within a radius at which the
galaxy may contain stars.
Nevertheless, given the Monte Carlo simulations discussed above, we conclude 
that this transient is likely associated only by chance with these galaxies.

For the remainder of sources, there are no obvious associations with galaxies
or stars.  The absence of optical or infrared counterparts only eliminates 
host candidates; the non-simultaneity of radio and optical observations 
implies that we place no limits on optical transients.

\section{Radio Transient Rates}

We can compute rates of transients in several different
ways.  We are limited in the accuracy of estimating transient rates 
by our lack of knowledge of the characteristic time scale, $t_{char}$, of
the transient sources.  This characteristic
time falls between the duration of the observations ($\sim 20$ min)
and the typical separation between epochs (7 days).  

The two-epoch survey sensitivity is independent of $t_{char}$.
The effective area for a two-epoch survey for a given flux-density limit is
just the sum of the areas in each epoch in which a source above 
the flux-density
limit is detectable.  For a set of $N_e$ images with uniform sensitivity over
a solid angle $\Omega$ in which $N_t$ transients are detected, 
the two-epoch source density is $N_t (N_e-1)^{-1} \Omega^{-1}$.  
For a sample with varying image properties,
we can compute the two-epoch source density 
given the image noise statistics, the synthesized beam size, 
and the primary beam shape.  The ratio of the primary beam solid angle to the 
synthesized beam solid angle gives the number of independent pixels in the 
image.  The source detection threshold per image is then set based on
the number of independent pixels and the rms noise.
We combine the 5 GHz and 8.4 GHz survey limits
and transient detections together.
The accumulated results as a function of flux-density threshold are 
listed in Table~\ref{tab:areasearched}.

We use a non-parametric 
Kaplan-Meier method \citep{1985ApJ...293..192F},
for estimating the cumulative source-count distribution
and then compute transient rates (Figure~\ref{fig:snapshotrate}).  
Seven transients are detected above a flux density of 370 $\mu$Jy, giving a
two-epoch source density of $1.5 \pm 0.4 {\rm\ deg^{-2}}$.  The 
transient rates 
are reasonably fit by a distribution with $S^{-\gamma}$ and $\gamma=1.5$, 
which corresponds to a non-evolving, volume-limited sample.  However,
the constraints on $\gamma$ are not very strong.  A larger $\gamma$ appears to
be required to achieve agreement between the rate determined from this
survey and the NVSS-FIRST survey
\citep{2006ApJ...639..331G}.  
Different event rates at 1.4 and 5 GHz due to synchrotron self-absorption
or other spectral index effects may also influence this comparison.
Additionally, the FIRST-NVSS survey may be incomplete due to confusion
from the mismatched resolution of these surveys.

We detected no transients in the 17 annual average images.  The typical
flux detection threshold at the half-power radius is 90 $\mu$Jy.  The total
two-epoch area surveyed at this sensitivity is 0.3 deg$^2$, implying a
$2\sigma$ upper
limit to the two-epoch density for year-long transients at 90 $\mu$Jy of 
$\sim 6\ {\rm deg^{-2}}$.  In 96 images composed of two-month averages, 
we found two transients.  At a flux-density threshold
of 200 $\mu$Jy, the effective area surveyed is 1.9 deg$^2$, giving a 
$2\sigma$ upper limit to the two-month transient rate 
of $\sim 2\ {\rm deg^{-2}}$.

We can also determine a rate as a function of area and time.  The
precision of this rate is limited by our lack of measurement of
the characteristic time of the transient events.  
The rate for sources brighter than 370 $\mu$Jy is then 
$0.07 < R < 40 {\rm\ deg^{-2} yr^{-1}}$.

If we assume that RTs 19840613 and 19870422 belong to a distinct 
transient class based on their association with blue galaxies, 
we can estimate a rate for such transients as a function of volume and time.  
We could detect this transient to a distance of $\sim 240$ Mpc, which gives
a total volume of $\sim 300$ Mpc$^3$.  Assuming $t_{char}=7 {\rm\ d}$,
we find a rate of $3 \times 10^5 {\rm\ Gpc^{-3} yr^{-1}}$.
Similarly for RT 19870422, we find a maximum volume of $\sim 5000$ Mpc$^3$
and use $t_{char}=60 {\rm\ d}$ to find a rate of
$4 \times 10^3 {\rm\ Gpc^{-3} yr^{-1}}$.  Jointly for RT 19840613 
and RT 19870422, the rate is $\sim 2 \pm 2 \times 10^5 {\rm\ Gpc^{-3} 
yr^{-1}}$.
These rates are consistent within an order of magnitude of the SN~Ib/c
rate determined optically, 
$4.8 \times 10^4  {\rm\ Gpc^{-3} yr^{-1}}$ 
\citep{2003ApJ...599..408B}.

There is some evidence that our VLA field contains an overdensity 
of galaxies at $z \approx 0.25$ (Table~\ref{tab:redshift}).  
Additionally, the cluster Abell 2047 is located within $9'$ of the
center of the VLA field \citep{1989ApJS...70....1A}.  A2047 is a low-density
cluster (richness class 0) with a radius $\sim 5'$ 
and a distance class of 6.  The distance class is an estimate of distance
based on visual inspection of photographic plates.  The distance class 
of 6 is consistent with a redshift of 0.25 for the cluster 
\citep{1987AJ.....93.1035S}.  If all 
the RTs are extragalactic, then the 
RT rates may be overestimated relative to the rate for the 
field by no more than a factor of a few.

\section{Discussion}

Two of the transients are reasonably identified as RSNe, possibly
Type Ib/c or II and similar to known RSNe associated with GRBs.
We consider now whether the eight transients not clearly
identified with nearby galaxies
are examples of known or expected
classes of radio transients.  These transients are unlikely to be RSNe
unless they are associated with faint and/or low surface brightness galaxies.
As we demonstrate below, the observed transient
parameters do not easily fit into any expected class.  

The known and anticipated sources of radio transients are very broad.  
Since we have no evidence for variability on a timescale of less than
20 min,
we exclude from further discussion transient sources that have characteristic 
timescales of less than 1~s.  These include pulsars and related
phenomena such as RRATS and giant pulses 
\citep{2006Nature...mcl,2003Natur.422..141H,2003ApJ...596.1142C},
as well as flares from 
extrasolar planets \citep{2000ApJ...545.1058B,2004ApJ...612..511L}.

\subsection{X-ray Sources}

Information on the X-ray properties of the transients and transient hosts
is very limited.
The ROSAT catalog of bright sources does not identify any sources
in the field of our survey \citep{1999A&A...349..389V}.
There are also no sources in the RXTE All Sky Monitor catalog in the field,
indicating an upper limit to the X-ray flux of events between 1997
and 2001 of $\sim 10$ mCrab \citep{1999NuPhS..69...12S}.
No observations of the field have been made with the Chandra 
or XMM Newton X-ray Observatories.

\subsection{Gamma-Ray Burst Afterglows}

Higher-energy counterparts to these transients have not been clearly 
detected. There are no known optically discovered SNe in the field of view
of our survey, nor have there been any optical variability campaigns in this
field.  There are no GRBs which are uniquely localized
to this field, although there are bursts with large positional uncertainties
that could have originated in the field of view of our survey.  We identify
7 BATSE events that occurred within $1\sigma$ of our field center between
1991 and 1999. Uncertainties in the BATSE positions, however, are 
$\sim 10^\circ$,
indicating a probability of $<0.1\%$ for even one of these GRBs to have 
occurred within our field
of view.  Of the GRBs, one event (GRB 990508) occurred 4~d after 
RT 19990504.  Given the low probability of concurrence and the
lack (in other GRBs) of radio emission before gamma-ray emission, 
we conclude that this is almost certainly a coincidence.  There is no 
systematic relationship between the remainder of the events and the 
radio transients.

\subsection{Orphan Gamma-Ray Burst Afterglows}

In the standard paradigm of GRBs and GRB afterglows, the gamma-ray emission is
highly anisotropic while the afterglow emission is less anisotropic or even 
isotropic. Thus, the observed GRBs and their afterglows from Earth are a 
subset of the total number of
GRBs. Afterglows detected without gamma-ray emission are known as orphan 
gamma-ray burst afterglows. The ratio of OGRBAs to GRBs is the 
inverse of the
beaming fraction, $f_b^{-1}$ \citep[e.g.,][]{2002ApJ...564..209D}. The 
phenomenon of a GRB afterglow without a corresponding gamma-ray burst has 
not been conclusively detected.

Detailed models of OGRBAs predict $\sim$1 per square degree at 
sensitivity thresholds
of 0.1 mJy, which is consistent with our number of transient detections 
\citep{2002ApJ...576..923L,2002ApJ...576..120T}.
Whereas the canonical radio afterglow  for the forward shock rises on a 
timescale of days and decays slowly, the majority of the radio 
transients found in the present survey occur on timescales 
less than 1 week.  One possible explanation in the context of 
GRB orphans is that the emission is due to the reverse shock, which is 
known, in the few cases where follow-up observations occurred 
soon after the GRB, to produce a bright and rapid ($<3$~d) radio signature 
\citep[e.g.,]{1999ApJ...522L..97K,2003astro.ph..9557F}.
Strong scintillation \citep[e.g.,]{1997Natur.389..261F}
of a standard, faint afterglow could also be responsible for producing 
short-term amplification that leads to an apparent fast transient, but then 
we might see marginal evidence for the forward shock in 
the epochs following the transient --- yet this is not seen 
in individual sources.

The single long-timescale transient (RT 20010331) could be consistent with 
having arisen from the long-lived forward shock (as it transitions from 
mildly relativistic to nonrelativistic).
Levinson et al. (2002) \citet{2002ApJ...576..923L}
calculate the expected number of forward-shock dominated OGRBAs for 
radio surveys. Taking the canonical interstellar medium density
of $n_0 = 1$ cm$^{-3}$, it is 
reasonable to expect that we only find 1 viable forward-shock candidate 
with the nominal beaming of $f_b^{-1} = 500$. Nevertheless, since for the 
parameters of our survey we probe only to $z \approx 0.6$, the 
non-detection of an apparent host galaxy in deep Keck imaging of
RT 20010331 would seem to suggest that our basic model assumptions 
for the radio afterglow are suspect. 

Thus, the transients that we have detected can be considered consistent 
with the expectations of OGRBAs.   There is, however, a large diversity 
of high-energy phenomena with poorly explored electromagnetic
signatures beyond the classical GRB afterglow picture, including
X-ray flashes, X-ray rich GRBs
\citep{2005ApJ...629..311S},
GRBs associated with SNe~Ib/c 
\citep{2003ApJ...599..408B}, and short-duration GRBs 
\citep{2005Natur.437..851G}.  The richness
of this phenomenology suggests that the radio transients we have discovered
could reflect the continuum of stellar collapse events.

\subsection{Active Galactic Nuclei}

AGNs are highly variable and common. 
With the appropriate selection technique, AGNs can be found with a
number density of $> 100$ per square degree 
\citep{2005ApJ...631..163S}.  If our
transients are AGNs, then the absence of deep, static detections indicates
they are associated with radio-quiet objects,
which constitute the majority of AGNs.  Short-timescale variability of
AGNs is often associated with interstellar scintillation and 
extreme scattering events, and typically has amplitudes of $\sim 10\%$ 
\citep{2003AJ....126.1699L}.
Variations of a factor of a few on timescales of hours 
have been seen in the most
extreme cases such as PKS 0405-385 and J1819+3845, 
which is not sufficient to account for the observed
transients given the very large amplitude of variation observed
\citep{2002ApJ...581..103R,2000ApJ...529L..65D}.

Intrinsic variations among luminous AGNs
are observed to have a 
timescale of months to years at radio frequencies 
\citep{1992ApJ...396..469H}.
Variations are rarely larger than a factor of a few in radio-loud objects.
A survey of variable background sources in the direction of M31 found
a small number of objects with variability of a factor of a few to 10 
on timescales of years \citep{2005ApJS..159..242G}.
An interesting example is the nucleus  of the
nearby spiral galaxy III~Zw~2, which has exhibited factor of 20 fluctuations
in its radio luminosity on timescales of months to years
\citep{2005A&A...435..497B}.
The TeV blazar Mrk 421 shows 50\% variations on
a timescale of $\sim 10$~d \citep{2005ApJ...630..130B}.
Sudden increases
in the accretion rate, or shocks in the jet,
could lead to a dramatic flare at radio wavelengths;
however, these flares are expected to have timescales on the order of
months to a year, inconsistent with our observations.

Recently discovered evidence for tidal disruption events in AGNs observed
in the ultraviolet indicate the possibility for large-amplitude AGN
transients \citep{2006IAUS..238E..97G}.  
However, the timescale for the observed UV transients
are months to a year rather than days.  Without a mechanism for a short
timescale in the radio, tidal disruption events are unlikely to account
for the observed radio transients.

\subsection{Stellar Sources}

We cannot exclude a Galactic origin for these transients.  A wide range of
stars and stellar systems are known to flare on timescales of minutes to 
hours to days.  However, no known stellar types closely fit the properties 
of these transients.  Our field ($b=36^\circ$, $l=115^\circ$)
is out of the plane of the Galaxy,
away from the Galactic center and bulge,
and is not near any known star cluster or star-forming region.
It is likely, therefore, that any stellar origin for these transients would
be even more numerous in one of these more favored directions.

\subsubsection{Known Luminous Transients}

The luminosity of the observed 
transients is $L_\nu \approx 10^{18} (d/{\rm 1\ kpc})^2
{\rm\ erg\ s^{-1}\ Hz^{-1}}$.  This is near the upper limit of 
the known stellar radio luminosity distribution 
\citep{2002ARA&A..40..217G}.
RS CVn binaries, FK Com class stars, Algol-class stars, and T Tauri
stars have been observed to radiate at this luminosity 
\citep[e.g.,][]{2003ApJ...598.1140B}.  
The total number of known RS CVn, FK Com stars, and Algol stars 
discovered at X-ray wavelengths is
$<1000$, indicating that we are unlikely to detect any such objects  by
chance, even if the distance cutoff in our survey is 10~kpc 
\citep{1993ApJS...86..599D}.
T Tauri stars are more common but are
unlikely to be found outside regions of active star formation.

\subsubsection{Late-Type Stars}

M-type dwarfs at distances as large as 1~kpc
have been identified as an important contribution to optical
transient rates \citep{Becker04,Kulkarni06}.
Due to the typical low radio luminosity 
($L_\nu \approx 10^{12}$ -- $10^{14} {\rm\ erg\ s^{-1}\ Hz^{-1}}$) of known
dMe stars and brown dwarfs, they are likely to be detected at 
distances of 30~pc or less.  In an extreme case, the dMe star
EV Lac was observed to undergo a flare of
a factor of $\sim 150$ to a maximum luminosity
$L_\nu \approx 10^{15} {\rm\ erg\ s^{-1}\ Hz^{-1}}$, 
with a rise time of minutes and a 
decay time of hours 
\citep{Osten06}.  Such a flare would be observed in our survey
to a distance of 100~pc.  
The local space density of stars with $M< 0.3~M_\sun$ is 
0.035 pc$^{-3}$, implying only a small chance of a low-mass star
or brown dwarf within the volume to which we are sensitive 
\citep{1999ApJ...521..613R}.
Any stellar origin for these transients must come from a faint
source, since we have no detection of stellar counterparts in Keck,
2MASS, or PAIRITEL images.  
The 2MASS magnitude limit of $K_s = 15$ implies that
any L dwarf ($M_K \approx 10$ mag) or earlier-type star must be at
a distance $> 100$~pc to remain undetected. PAIRITEL limits require 
such stars to be at a distance $>500$ pc.  
The limiting Keck magnitude requires late-type M stars to 
be at a distance $>1000$ pc.

The probability distribution of radio activity in low-mass stars 
is not well quantified.  Few stars have been observed at radio 
wavelengths for more than a few hours or a few days.  
Despite the lack of long-term monitoring of dMe
stars, one can estimate the probability distribution
of flares by comparison with the properties of the Sun.  The 
probability of a solar flare at centimeter wavelengths
of a given flux density $S$ scales as $S^{-1.7}$ \citep{Nita04}.  
>From observations of a large sample of late-type stars
\citep{Berger06}, we estimate a probability of $\sim 0.01$ 
for a late-type star to exhibit a  luminosity
$L_\nu \approx 10^{15} {\rm\ erg\ s^{-1}\ Hz^{-1}}$.  
If the probability of stellar flares scales with the same relation as
solar flares, a flare of $L_\nu \approx 10^{18}
{\rm\ erg\ s^{-1}\ Hz^{-1}}$ from a late-type star at a distance of 
1~kpc would be visible in our radio survey but 
missing in the Keck image, and it would occur with a probability
of $<10^{-7}$.  Assuming a constant density of stars set to the local 
value, the total number of M dwarfs out to 1~kpc within our field
of view is $<5 \times 10^3$.  Given our $\sim 1000$ observations, the expected
number of flares of this magnitude detected is $\lsim 1$.  This is an upper
limit since the statistics for the observed stars are biased by selection
for activity at other wavelengths.  Within an order of
magnitude, though, this matches the number of transients that we detect, 
suggesting that
low-mass stars at 1~kpc or greater distance may contribute to, but are
unlikely to dominate, the observed radio-transient rate.

Very large radio flares might be accompanied by luminous X-ray flares.  
If the transients follow the radio/X-ray correlation 
\citep{2002ARA&A..40..217G}, then
a radio flare of $10^{18} {\rm\ erg\ s^{-1}\ Hz^{-1}}$ at a distance
of 1~kpc would have an X-ray luminosity of 
$10^{-9 \pm 1}{\rm\ erg\ s^{-1}}$, which exceeds the RXTE ASM detection 
threshold of $\sim  10^{-10} {\rm\ erg\ s^{-1}}$.  However, many sources
do not closely follow this relationship, including many low-mass
stars \citep[e.g.,][]{2001Natur.410..338B}.

\subsubsection{Soft Gamma-Ray Repeaters}

The transients share some properties with 
soft gamma-ray repeaters (SGRs).  SGRs are magnetars that exhibit 
dramatic flares at gamma-ray and radio wavelengths
\citep[e.g.,][]{1999Natur.398..127F,2005Natur.434.1104G}.  
SGR~1806--20 produced a flare that reached a luminosity of 
$5 \times 10^{22} {\rm\ erg\ s^{-1}\ Hz^{-1}}$, had a characteristic
timescale of days, and declined as $t^{-2.7}$.  Such a flare would be
detectable in our dataset at a distance of $\sim 200$ kpc, would
appear in only a single epoch, would fade away rapidly so that it is not
seen in long-term averages, and would have no lasting optical counterpart.  
If all of our transients are Milky Way
SGRs to a distance of $\lsim 200$ kpc, the implied rate is 
$\gsim 0.01 {\rm\ kpc^{-3}\ y^{-1}}$ at high
Galactic latitude.  Without making any correction for Galactic population
distribution, we estimate a total rate of 
$\gsim 3 {\rm\ yr^{-1}}$ in the disk of the Galaxy,
which is at least an order of magnitude too high relative to the number
of discovered SGRs.

\subsubsection{X-ray Binaries}

X-ray binaries (XRBs) can be seen throughout the Galaxy 
\citep[e.g.,][]{2005ApJ...633..218B} but are rare
\citep{2005ApJ...622L.113M}.  For the extremely overdense Galactic center,
the XRB detection rate in the radio is $\lsim 0.01 {\rm\ kpc^{-3}\ yr^{-1}}$.
The comparable observed transient rate would imply a similar overdensity
at high Galactic latitude, which is implausible.

\subsubsection{Pulsars}

Most pulsar variability occurs on time scales much shorter than our 
observations and is therefore unlikely to account for our observed 
transients.  There have been some examples, however, of longer timescale
variability.  PSR B1931+24, for instance, has periods of
activity and inactivity with characteristic times of days to tens of
days \citep{2006Sci...312..549K}.  This variability is quasi-periodic, however, indicating
that we would detect it and similar pulsars in multiple epochs throughout
our survey.

\subsection{Propagation Effects}

We have already referred to the phenomenon of intra-day variability (IDV), 
which in the most extreme cases leads to variations of a factors of a 
few on a timescale of hours.  IDV of this scale has been demonstrated to 
originate from interstellar scintillation of compact structure in the
source.  \citet{1997A&A...325..631G} 
identified a factor of 43 fluctuation in the flux density 
of PSR B0655+64 at 325~MHz, which they argue originates from strong
focusing in the interstellar medium --- that is, a caustic.  It is difficult
to estimate the frequency with which such a phenomenon can occur, but 
observational evidence is scarce.

Microlensing could also produce a transient amplification of a background
source.   A simple calculation estimates that to explain all 10 RTs as 
the result of microlensing requires the product
$n\tau \lsim 10^{-2}$, where $n$ is the number of sources that can be lensed
and $\tau$ is the lensing opacity \citep{1996ARA&A..34..419P}.  
Since $\tau$ has been measured
to be $\sim 10^{-7}$ by MACHO and OGLE experiments, then $n \gsim 10^9$, 
which is absurdly large for any known population of Galactic or extragalactic
radio sources.

Finally, reflected solar flares off objects in the Solar System could 
produce detectable radio flares.  A 1~MJy solar flare reflecting off a 
1000~km object at a distance of 1~AU from the Earth would produce a 
0.1~mJy flare at Earth.  Our observed field is far out of the 
ecliptic plane, however, and there are few known Solar-System objects 
of this size outside of the ecliptic. Nearer objects could be smaller 
but would move significantly during the course of the observation.

\section{Conclusions}

We have conducted a 944-epoch, 20-yr-long survey for radio transients
with archival 5 and 8.4 GHz Very Large Array data.  Ten radio transients
are apparent in this data set. Eight transients appear in only a single epoch
and disappear completely thereafter.  Two transients appear in two-month
integrations and also never reappear.  We estimate the rate of radio
transients below 1~mJy with a large uncertainty due to the wide range
of characteristic times possible for these sources.

Two of the transients may be peculiar 
Type Ib/c and/or Type II radio SNe in nearby galaxies.  
All of these transients share some characteristics with those expected
of orphan gamma-ray burst afterglows, but they cannot be conclusively
identified as such.  The diversity of high-energy stellar-death
phenomena makes it difficult to
place specific limits on the nature of GRBs based on these results.
The absence of persistent counterparts to these sources indicate that
they are unlikely to be AGNs.
Galactic sources appear unlikely to be an explanation, but
they certainly cannot be ruled out.  In particular, late-type stars could
contribute to the number of radio transients if their luminosity distribution
can be extrapolated by several orders of magnitude beyond the most
luminous observed M-dwarf flare.
Deeper multi-wavelength imaging is critical for identifying the hosts
of these events.

In the near future, the Allen Telescope Array \citep{2004SPIE.5489.1021D}
will be efficient at
discovering transients of this sort.  The wide field of view of this
telescope makes it a powerful instrument for large solid angle radio
surveys.  Given the apparently short timescale of these
phenomena, rapid response to transient discovery at radio and other
wavelengths will be critical for determining the nature of these 
transients.

\acknowledgements

These observations made use of archival data from the Very Large Array.
The National Radio Astronomy Observatory is a facility of the National 
Science Foundation (NSF) operated under cooperative agreement by 
Associated Universities, Inc.  This research utilized the NASA/IPAC 
Extragalactic Database (NED), which is operated by the Jet 
Propulsion Laboratory, California Institute of Technology, under 
contract with the National Aeronautics and Space Administration (NASA). 
It also used the SIMBAD database, operated at CDS, 
Strasbourg, France, as well as data from the University of Michigan 
Radio Astronomy Observatory, which has been supported by the 
University of Michigan and the NSF.
Some of the data presented here were obtained at the W. M. Keck
Observatory, which is operated as a scientific partnership among the
California Institute of Technology, the University of California,
and the National Aeronautics and Space Administration. The Observatory
was made possible by the generous financial support of the W. M. Keck
Foundation. We thank the Keck staff for their assistance.
The Peters Automated Infrared 
Imaging Telescope (PAIRITEL) is operated by the Smithsonian 
Astrophysical Observatory (SAO) and was made possible by a grant 
from the Harvard University Milton Fund, a camera loan from the 
University of Virginia, and the continued support of the SAO,
M. Skrutskie, and UC Berkeley. The PAIRITEL project and D.P. are 
further supported by NASA/Swift Guest Investigator Grant NNG06GH50G.
A.V.F. acknowledges NSF grant AST--0607485 and NASA/Swift grant 
NNG06GI86G. We thank Daniel Stern, Mark Dickinson, and Hyron Spinrad 
for assistance with some of the Keck observations and analysis.


\begin{deluxetable}{rrrrr}
\tabletypesize{\scriptsize}
\tablecaption{Steady Sources Detected in at Least One Epoch}
\tablehead{\colhead{RA} & \colhead{Dec.} & \colhead{$N_{det}$} &
\colhead{$<S>$} & \colhead{$S_{deep}$} \\
\colhead{(J2000)} & \colhead{(J2000)} &                              &
\colhead{($\mu$Jy)} & \colhead{($\mu$Jy)}  
\label{tab:steady}}
\startdata
\multicolumn{5}{c}{5 GHz} \\
15 00 11.48 $\pm$ 0.10 & 78 12 42.26 $\pm$ 00.13 & 26 & 2380 $\pm$ 73 & 502 $\pm$ 17 \\
15 00 44.87 $\pm$ 0.10 & 78 18 39.04 $\pm$ 00.32 & 4 & 874 $\pm$ 36 & 180 $\pm$ 7 \\
15 01 11.70 $\pm$ 0.10 & 78 15 20.23 $\pm$ 00.10 & 141 & 504 $\pm$ 6 & 307 $\pm$ 4 \\
15 01 16.19 $\pm$ 0.10 & 78 12 45.88 $\pm$ 00.10 & 363 & 1146 $\pm$ 14 & 741 $\pm$ 5 \\
15 01 22.67 $\pm$ 0.10 & 78 18 05.66 $\pm$ 00.10 & 452 & 880 $\pm$ 9 & 634 $\pm$ 3 \\
15 02 51.40 $\pm$ 0.10 & 78 18 12.46 $\pm$ 00.10 & 23 & 351 $\pm$ 19 & 155 $\pm$ 3 \\
15 03 24.83 $\pm$ 0.14 & 78 17 37.62 $\pm$ 00.46 & 1 & 318 $\pm$ 54 & 166 $\pm$ 4 \\
15 03 28.07 $\pm$ 0.30 & 78 09 21.50 $\pm$ 01.04 & 1 & 2045 $\pm$ 301 & $2650 \pm 8 $ \\
\multicolumn{5}{c}{8.4 GHz} \\
15 01 11.13 $\pm$ 0.12 & 78 15 22.45 $\pm$ 00.32 & 1 &1081 $\pm$ 154 & 291 $\pm$ 9 \\
15 01 22.58 $\pm$ 0.10 & 78 18 05.80 $\pm$ 00.10 & 159 & 1099 $\pm$ 36 & 794 $\pm$ 8 \\
15 02 51.38 $\pm$ 0.10 & 78 18 13.14 $\pm$ 00.18 & 3 & 518 $\pm$ 31 & 225 $\pm$ 4 \\
15 03 23.41 $\pm$ 0.41 & 78 17 35.74 $\pm$ 00.64 & 1 & 840 $\pm$ 157 & 212 $\pm$ 8 \\
\enddata
\tablecomments{Columns are (1) right ascension, (2) declination, (3)
the number of epochs in which the source is detected,
(4) mean flux density of detections, and (5) flux density in the deep image.}
\end{deluxetable}

\begin{deluxetable}{rrrrrrrrr}
\tabletypesize{\scriptsize}
\tablecaption{Transient Sources from a Single Epoch}
\tablehead{\colhead{Epoch} &\colhead{RA} & \colhead{Dec.} & 
\colhead{$S$} & \colhead{$S_{deep}$} & \colhead{$t_{next}$} & \colhead{$S_{next}$} & \colhead{$-\log_{10}{PFD}$} & \colhead{Host} \\
 &  \colhead{(J2000)} & \colhead{(J2000)} &
\colhead{($\mu$Jy)} & \colhead{($\mu$Jy)} & \colhead{(days)} & \colhead{($\mu$Jy)} 
\label{tab:transients}}
\startdata
\multicolumn{8}{c}{5 GHz} \\
19840502 & 15 02 24.61 $\pm$ 0.21 & 78 16 10.08 $\pm$ 00.46 & 448 $\pm$ 74 & $< 8$ & 7 & $ -10 \pm 68 $& 4.5 & X\\
19840613 & 15 01 38.07 $\pm$ 0.21 & 78 18 40.75 $\pm$ 00.39 & 566 $\pm$ 81 & $< 28$ & 7 & $ 86 \pm 89 $& 7.5 & G\\
19860115 & 15 02 26.40 $\pm$ 0.44 & 78 17 32.39 $\pm$ 01.59 & 370 $\pm$ 67 & $< 8$ & 7 & $ 199 \pm 121 $& 3.6 & X \\
19860122 & 15 00 50.15 $\pm$ 0.34 & 78 15 39.37 $\pm$ 01.38 & 1586 $\pm$ 248 & $< 15$ & 7 & $ -59 \pm 164 $& 6.4 & X\\
19920826 & 15 02 59.89 $\pm$ 0.35 & 78 16 10.82 $\pm$ 02.44 & 642 $\pm$ 101 & $< 9$ & 56 & $ 37 \pm 83 $& 6.3 & ?\\
19970528 & 15 00 23.55 $\pm$ 0.10 & 78 13 01.37 $\pm$ 00.17 & 1731 $\pm$ 232 & $< 36$ & 7 & $ 90 \pm 206 $& 7.9 & ? \\
19990504 & 14 59 46.42 $\pm$ 0.56 & 78 20 29.03 $\pm$ 00.74 & 7042 $\pm$ 963 & $< 117$ & 21 & $ -313 \pm 1020 $& 9.1 & X\\
\multicolumn{8}{c}{8.4 GHz} \\
19970205 & 15 01 29.35 $\pm$ 0.10 & 78 19 49.20 $\pm$ 00.10 & 2234 $\pm$ 288 & $<646$ & 5 & $857 \pm 323$ & 8.0 & X\\
\enddata
\tablecomments{Columns are (1) RT epoch, (2) right ascension, (3) declination, 
(4) mean flux density of detection, (5) flux density in the deep image, 
(6) separation in days between next epoch and epoch of transient, 
(7) flux density in the next epoch, 
(8) the negative logarithm of the probability of false detection in 
the single epoch, and (9) optical/IR host identification, where ``G'' 
indicates galaxy, ``?'' indicates possible galaxy association, and ``X'' 
indicates no host.}
\end{deluxetable}

\begin{deluxetable}{rrrrrrrrr}
\tabletypesize{\scriptsize}
\tablecaption{Transient Sources with $T_{\rm char}=2$ months}
\tablehead{\colhead{Epoch} &\colhead{RA} & \colhead{Dec.} & 
\colhead{$S$} & \colhead{$S_{deep}$} & \colhead{$t_{next}$} & \colhead{$S_{next}$} & \colhead{$-\log_{10}{PFD}$} & \colhead{Host} \\
 &  \colhead{(J2000)} & \colhead{(J2000)} &
\colhead{($\mu$Jy)} & \colhead{($\mu$Jy)} & \colhead{(days)} & \colhead{($\mu$Jy)} 
\label{tab:transientsmonths}}
\startdata
\multicolumn{8}{c}{5 GHz} \\
19870422 & 15 00 50.01 $\pm$ 0.35 & 78 09 45.49 $\pm$ 0.97 & $505 \pm 83$ & $<36$ & 96 & $-154 \pm 65$ & 3.0 & G\\
\multicolumn{8}{c}{8.4 GHz} \\
20010331 & 15 03 46.18 $\pm$ 0.10 & 78 15 41.68 $\pm$ 00.11 & 697 $\pm$ 94 & $<37$ & 59 & 85 $\pm$ 85 & 3.4 & X\\
\enddata
\tablecomments{Columns are (1) RT epoch, (2) right ascension, (3) declination, 
(4) mean flux density of detection, (5) flux density in the deep image, 
(6) separation in days between next epoch and epoch of transient, 
(7) flux density in the next epoch, (8) the negative logarithm of 
the probability of false detection in the single epoch, and (9) 
optical/IR host identification, where ``G'' indicates galaxy, 
``?'' indicates possible galaxy association, and ``X'' indicates no host.}
\end{deluxetable}

\begin{deluxetable}{lrrrrr}
\tablecaption{Magnitudes, Colors, and Redshifts of Transient Host Galaxies}
\tablehead{\colhead{RT Epoch} & \colhead{MAPS Counterpart} & \colhead{Host Offset} & \colhead{$R$} & \colhead{$B-R$} & \colhead{$z$}  \\
                              &                &                  & \colhead{(mag)} & \colhead{(mag)} 
\label{tab:redshift}}
\startdata
19840613 & P023-0189928 & 2.7$^{\prime\prime}$ NW & 16.5 & 1.8 & 0.040 \\
19860115-1 & P023-0190130 & 15$^{\prime\prime}$ NE & 19.2 & 1.3 & 0.130 \\
19860115-2 & \dots        & 20$^{\prime\prime}$ SW &      &     & 0.242 \\
19870422 & P023-0189613 & 1.5$^{\prime\prime}$ SE & 20.2 & 2.5 & 0.249 \\
19920826 & \dots        & 5$^{\prime\prime}$ NW &        &     & unknown \\
19970528 & P023-0189499 & 5$^{\prime\prime}$ SE & 19.6   & 2.4 & 0.245 \\
\enddata
\tablecomments{Host offset is position relative to the transient.
Magnitudes and colors are from MAPS catalog.}
\end{deluxetable}

\begin{deluxetable}{rrrrrrr}
\tabletypesize{\scriptsize}
\tablecaption{Effective Area Searched for a Comparable Two-Epoch Survey}
\tablehead{            & \multicolumn{2}{c}{5 GHz} & \multicolumn{2}{c}{8.4 GHz} & Total \\
\colhead{Flux Density} & \colhead{Number of Epochs} & \colhead{Area} & \colhead{Number of Epochs} & \colhead{Area} & \colhead{Number of Epochs} & \colhead{Area} \\
\colhead{($\mu$Jy)} &  & \colhead{(deg$^2$)}&  & \colhead{(deg$^2$)}&  & \colhead{(deg$^2$)} 
\label{tab:areasearched}}
\startdata
  70 &    9 &  0.16 &	0	&	0.0 &	9 & 0.16\\ 
 100 &   10 &  0.24 &	0	&	0.0 &	10 & 0.24\\ 
 140 &   12 &  0.34 &	0	&	0.0 &	12 & 0.34\\ 
 200 &   35 &  0.50 &	0	&	0.0 &	35 & 0.50\\ 
 280 &  201 &  1.42 &	442	&	1.45&	435 & 2.19\\ 
 400 &  464 &  4.57 &	560	&	3.08&	761 & 6.20\\ 
 560 &  561 &  9.14 &	584	&	4.79&  873 & 11.68\\ 
 800 &  596 & 14.56 &	591	&	6.64&  910 & 18.09\\ 
1120 &  606 & 19.85 &	591	&	8.40&  924 & 24.31\\ 
1600 &  607 & 25.51 &	598	&      10.28&  925 & 30.97\\ 
\enddata
\end{deluxetable}

\begin{deluxetable}{rrrrrrr}
\tabletypesize{\scriptsize}
\tablecaption{Effective Area Searched for a Comparable Two-Epoch Survey for $T_{\rm char}=2$ months}
\tablehead{            & \multicolumn{2}{c}{5 GHz} & \multicolumn{2}{c}{8.4 GHz} & Total \\
\colhead{Flux Density} & \colhead{Number of Epochs} & \colhead{Area} & \colhead{Number of Epochs} & \colhead{Area} & \colhead{Number of Epochs} & \colhead{Area} \\
\colhead{($\mu$Jy)} &  & \colhead{(deg$^2$)}&  & \colhead{(deg$^2$)}&  & \colhead{(deg$^2$)} 
\label{tab:areasearchedmonths}}
\startdata
  70 &   24 &  0.10 &    9 &  0.02 &  26  &  0.1  \\
 100 &   41 &  0.49 &   38 &  0.22 &  54  &  0.56  \\
 140 &   55 &  0.91 &   61 &  0.67 &  73  &  1.11  \\
 200 &   75 &  1.51 &   74 &  1.31 &  99  &  1.91  \\
 280 &   90 &  2.24 &   78 &  1.97 &  115  &  2.84  \\
 400 &   95 &  3.11 &   78 &  2.69 &  120  &  3.94  \\
 560 &   95 &  3.93 &   78 &  3.37 &  120  &  4.97  \\
 800 &   95 &  4.81 &   78 &  4.09 &  120  &  6.07  \\
1120 &   95 &  5.64 &   78 &  4.77 &  120  &  7.11  \\
1600 &   95 &  6.52 &   78 &  5.48 &  120  &  8.22  \\
\enddata
\end{deluxetable}

\begin{deluxetable}{rrrrrrr}
\tabletypesize{\scriptsize}
\tablecaption{Effective Area Searched  for a Comparable Two-Epoch Survey for $T_{\rm char}=1$ year}
\tablehead{            & \multicolumn{2}{c}{5 GHz} & \multicolumn{2}{c}{8.4 GHz} & Total \\
\colhead{Flux Density} & \colhead{Number of Epochs} & \colhead{Area} & \colhead{Number of Epochs} & \colhead{Area} & \colhead{Number of Epochs} & \colhead{Area} \\
\colhead{($\mu$Jy)} &  & \colhead{(deg$^2$)}&  & \colhead{(deg$^2$)}&  & \colhead{(deg$^2$)} 
\label{tab:areasearchedyears}}
\startdata
  50 &    5 &  0.02 & 10 & 0.02 & 15 & 0.04\\
  70 &   13 &  0.09 & 14 & 0.05 & 15 & 0.10\\
 100 &   17 &  0.24 & 15 & 0.08 & 20 & 0.26\\
 140 &   17 &  0.38 & 15 & 0.14 & 20 & 0.42\\
 200 &   17 &  0.53 & 16 & 0.18 & 21 & 0.57\\
 280 &   17 &  0.67 & 16 & 0.23 & 21 & 0.73\\
 400 &   17 &  0.82 & 16 & 0.27 & 21 & 0.89\\
 560 &   17 &  0.96 & 16 & 0.32 & 21 & 1.14\\
 800 &   17 &  1.09 & 16 & 0.35 & 21 & 1.19\\
1120 &   17 &  1.15 & 16 & 0.36 & 21 & 1.24\\
1600 &   17 &  1.16 & 16 & 0.36 & 21 & 1.25\\
\enddata
\end{deluxetable}

\includegraphics[width=3in,angle=-90]{f1a.ps}\includegraphics[width=3in,angle=-90]{f1b.ps}
\figcaption{Image from epochs 19840613 (left) and 19840620 (right) at 5~GHz.  
The two westernmost sources are steady
sources J150112+781520 and J150123+781806. The northernmost source is a
transient, J150138+781841, detectable only in epoch 19840613. 
Contours are 5, 6, and 7 times the image rms of 57 and 63 $\mu$Jy.
\label{fig:typicalimage}}

\plotone{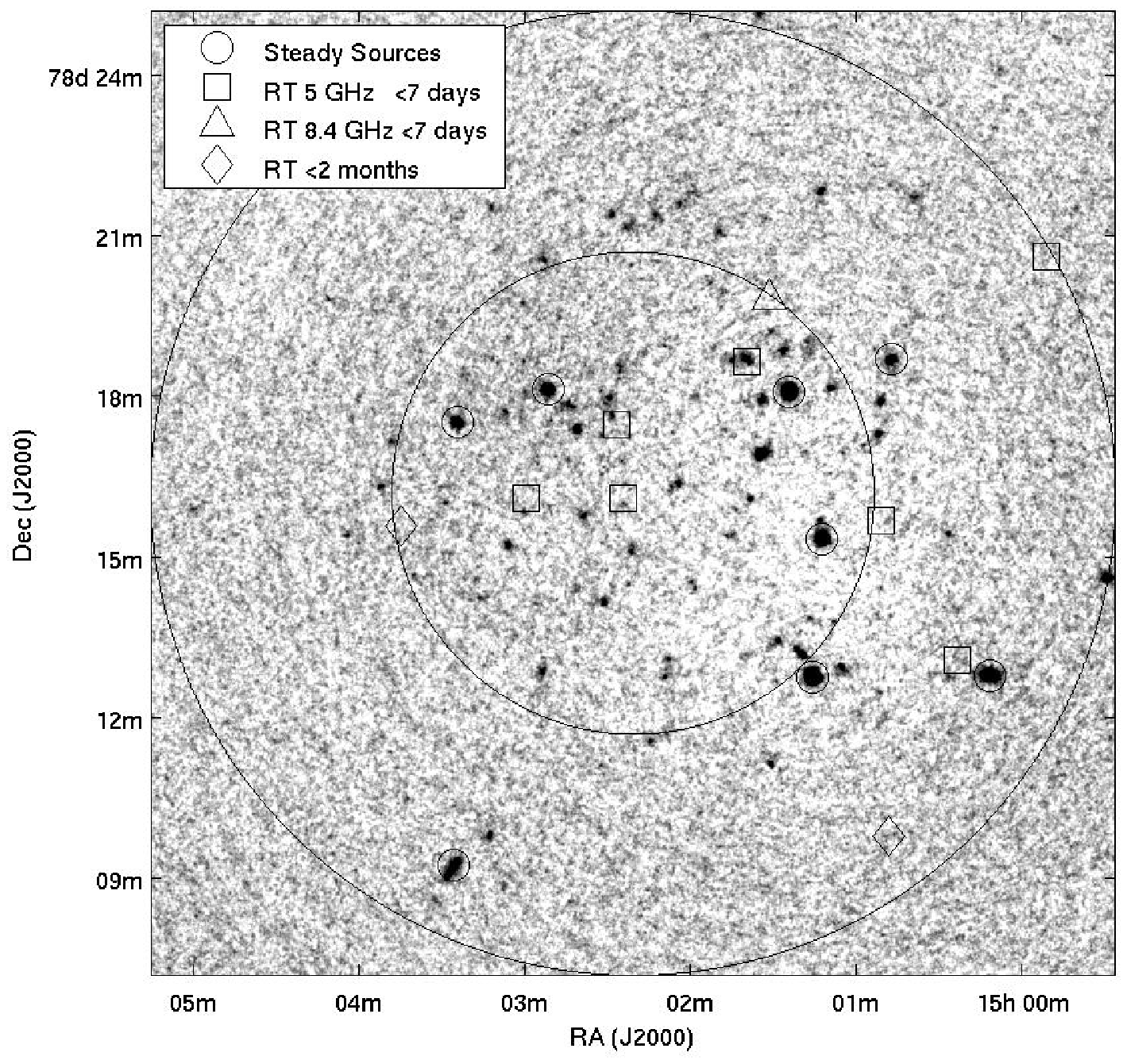}
\figcaption[]{Deep image obtained at 5~GHz.  We show the positions 
of single-epoch transient sources at 5~GHz (squares), single-epoch 
transient source at 8.4 GHz (triangle), two-month epoch transient
sources (diamonds), and steady sources detected at least once in 
an individual epoch (circles). The two large circles indicate the 
half-power radius and twice the half-power radius of the 5-GHz field.
The grey scale stretches from $-$2.5 to 30 $\mu$Jy.  The image has not been
corrected for primary beam attenuation.
\label{fig:deep}}

\plotone{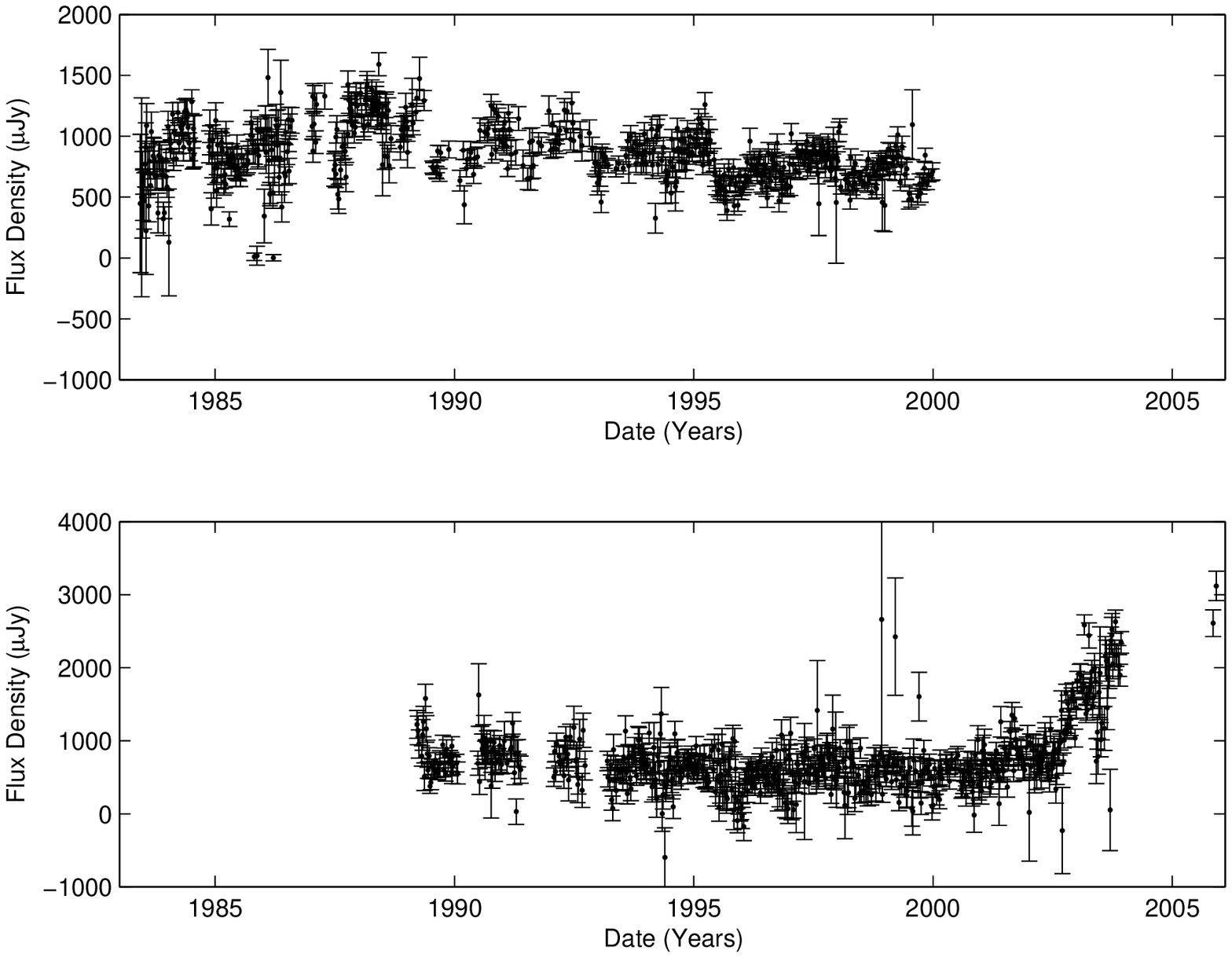}
\figcaption[]{Light curve of J150123+781806 at 5 GHz (top) and 
8.4 GHz (bottom) from the entire data set.
\label{fig:standardlightcurve}}

\mbox{\includegraphics[width=3in]{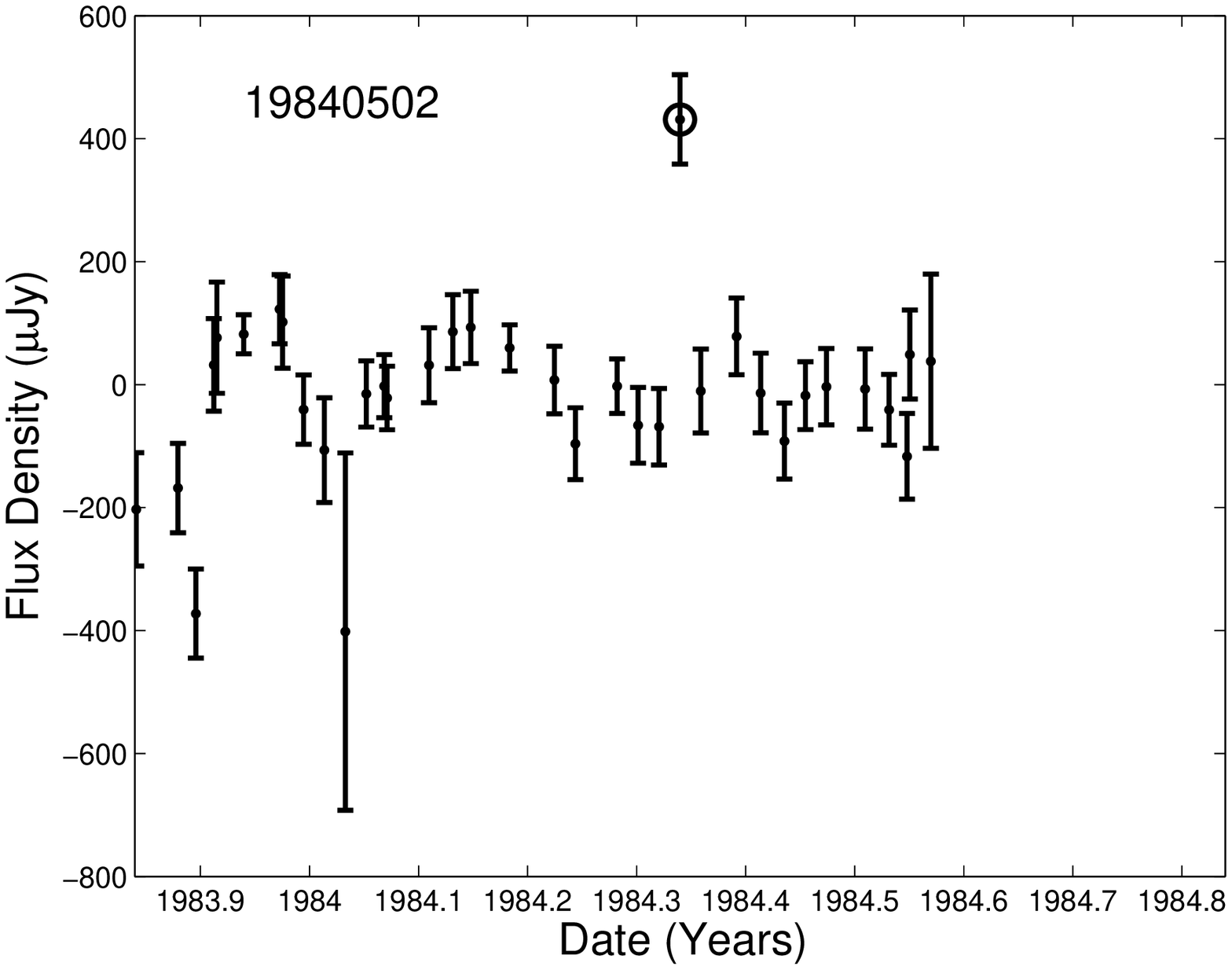}\includegraphics[width=3in]{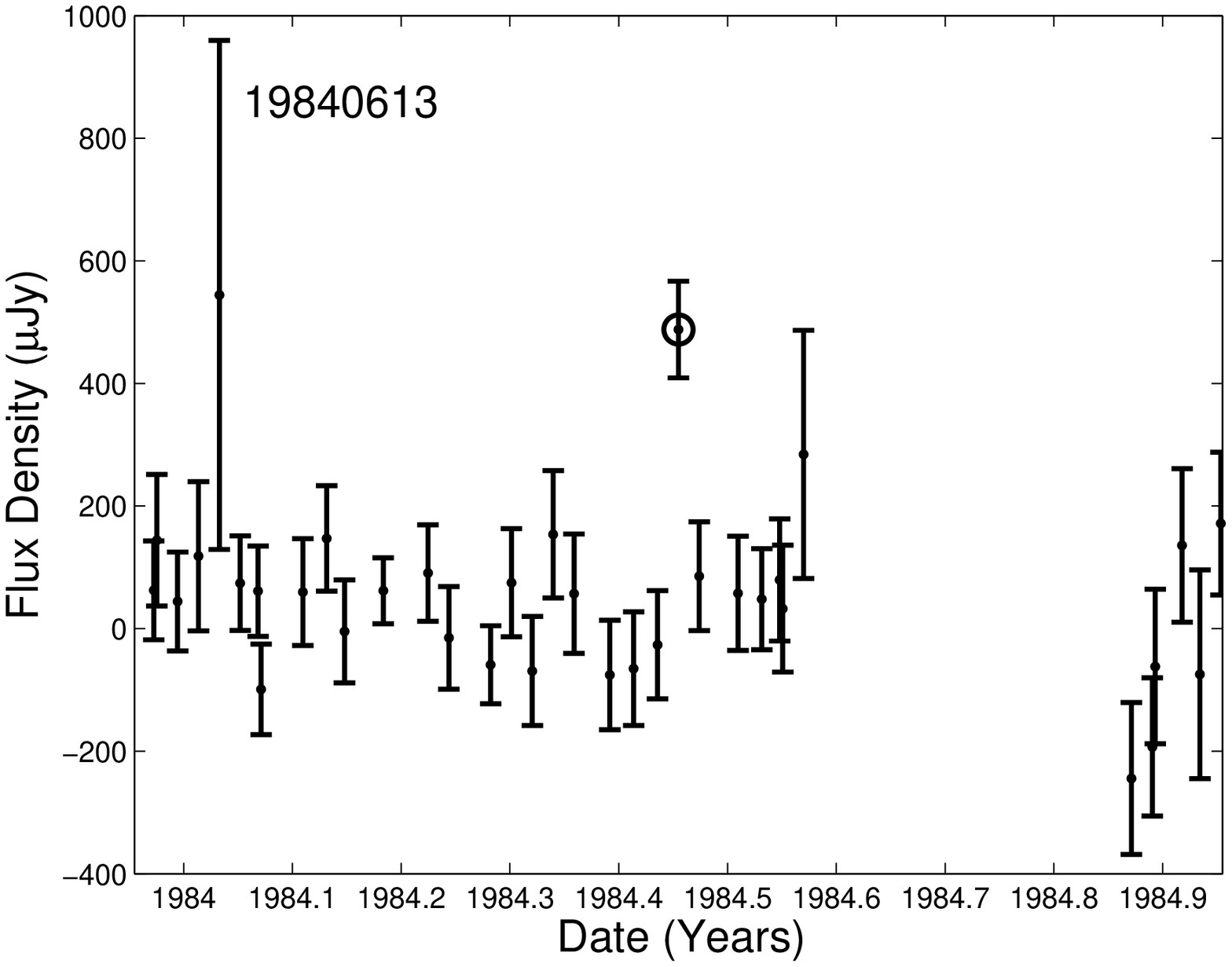}}
\mbox{\includegraphics[width=3in]{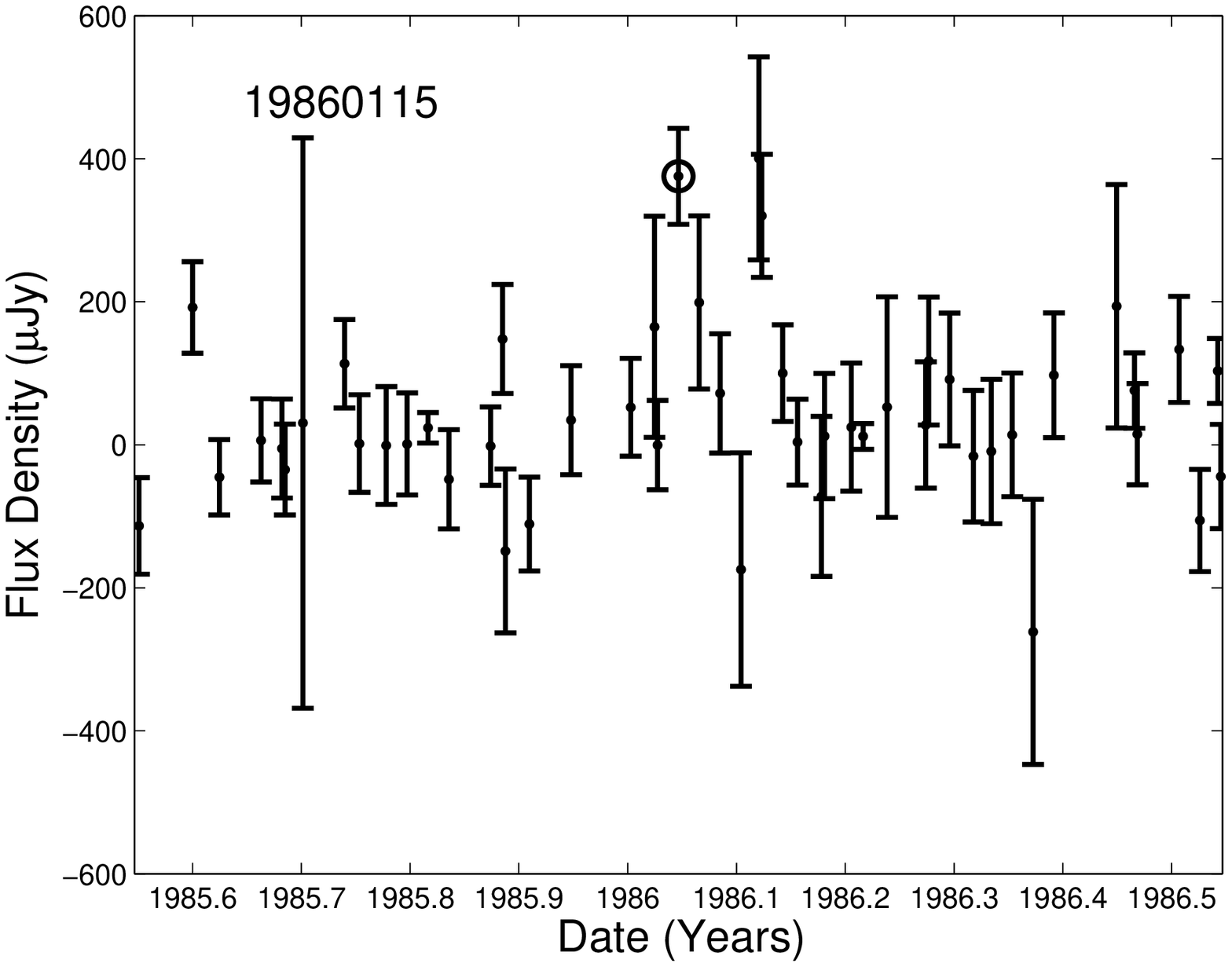}\includegraphics[width=3in]{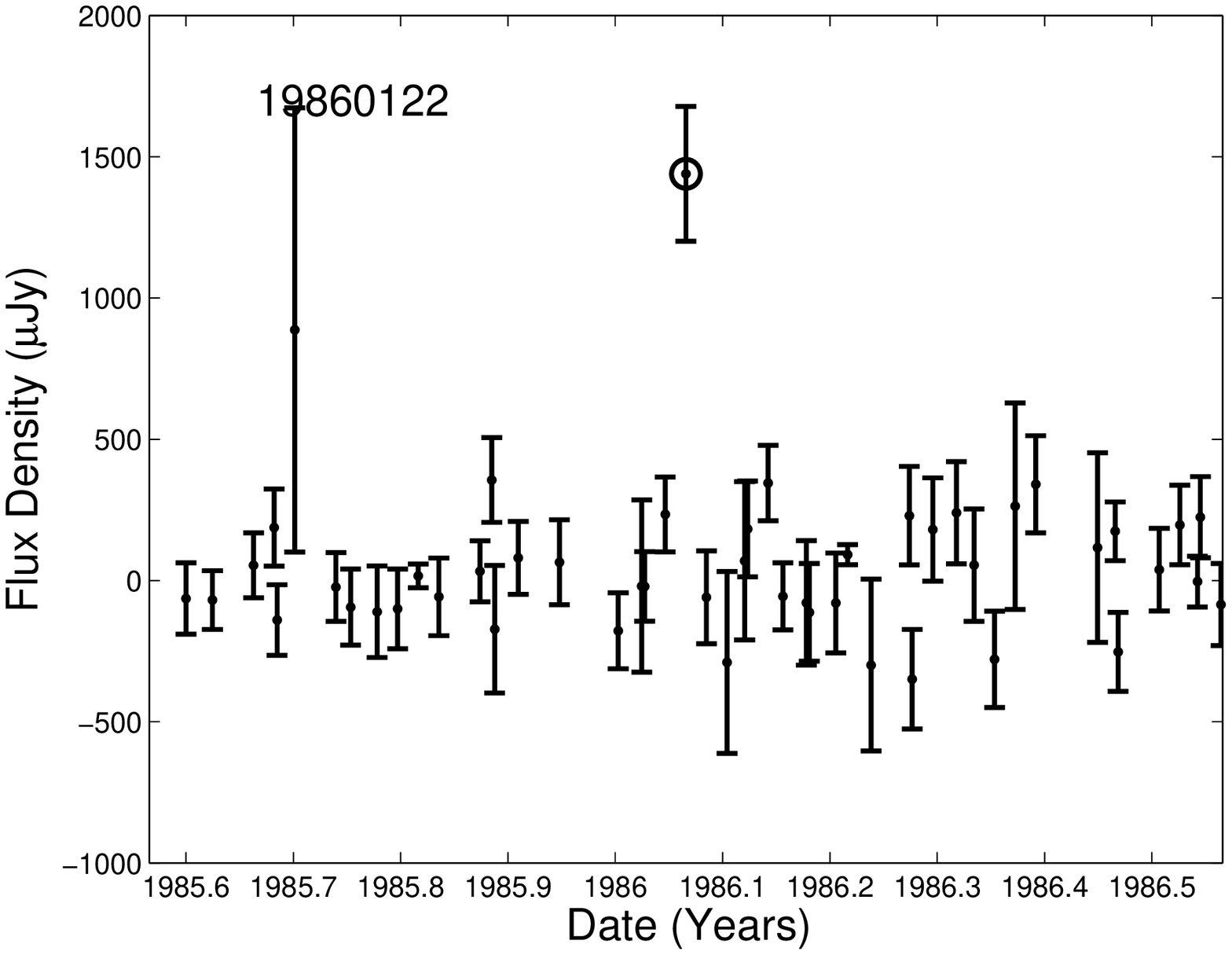}}
\mbox{\includegraphics[width=3in]{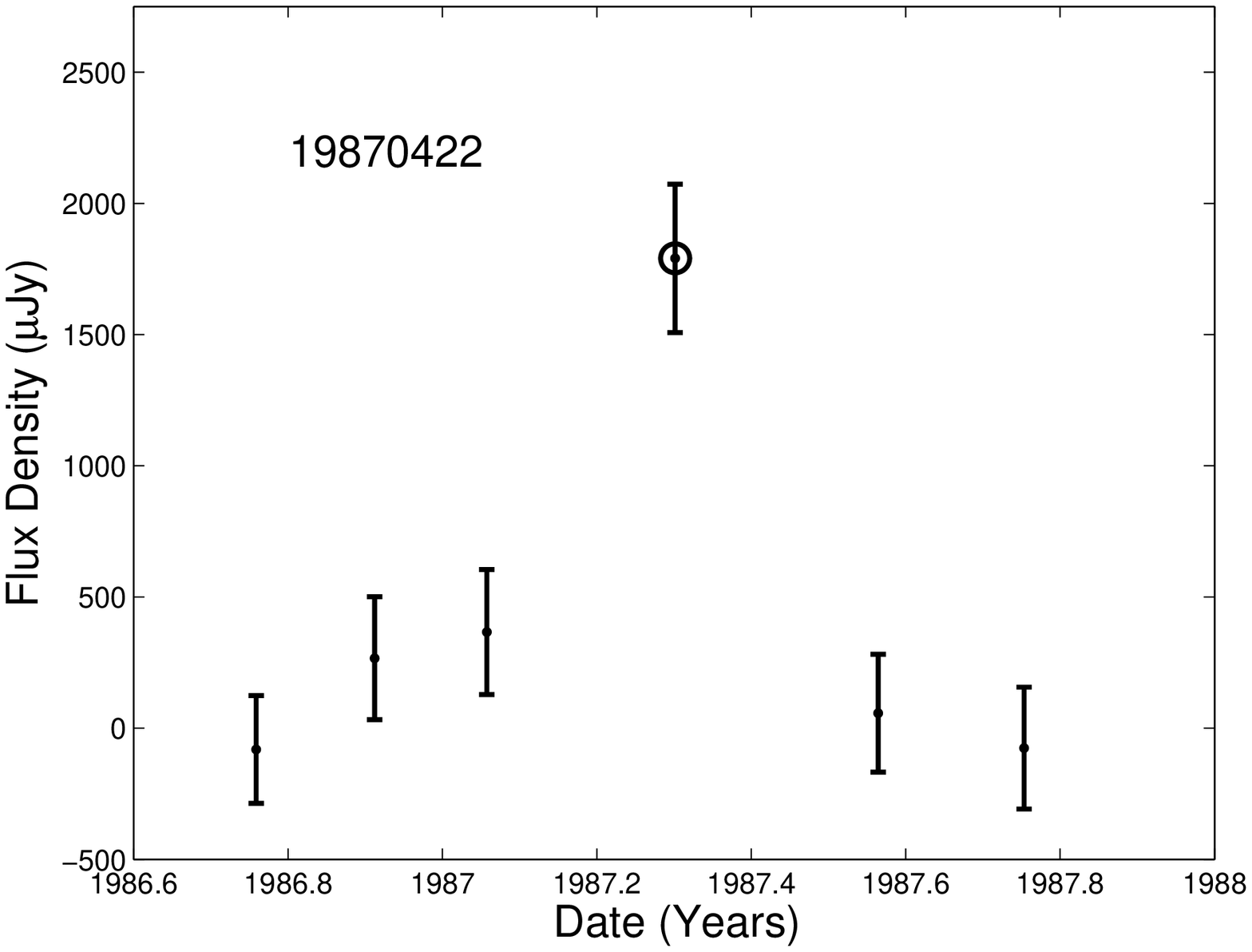}\includegraphics[width=3in]{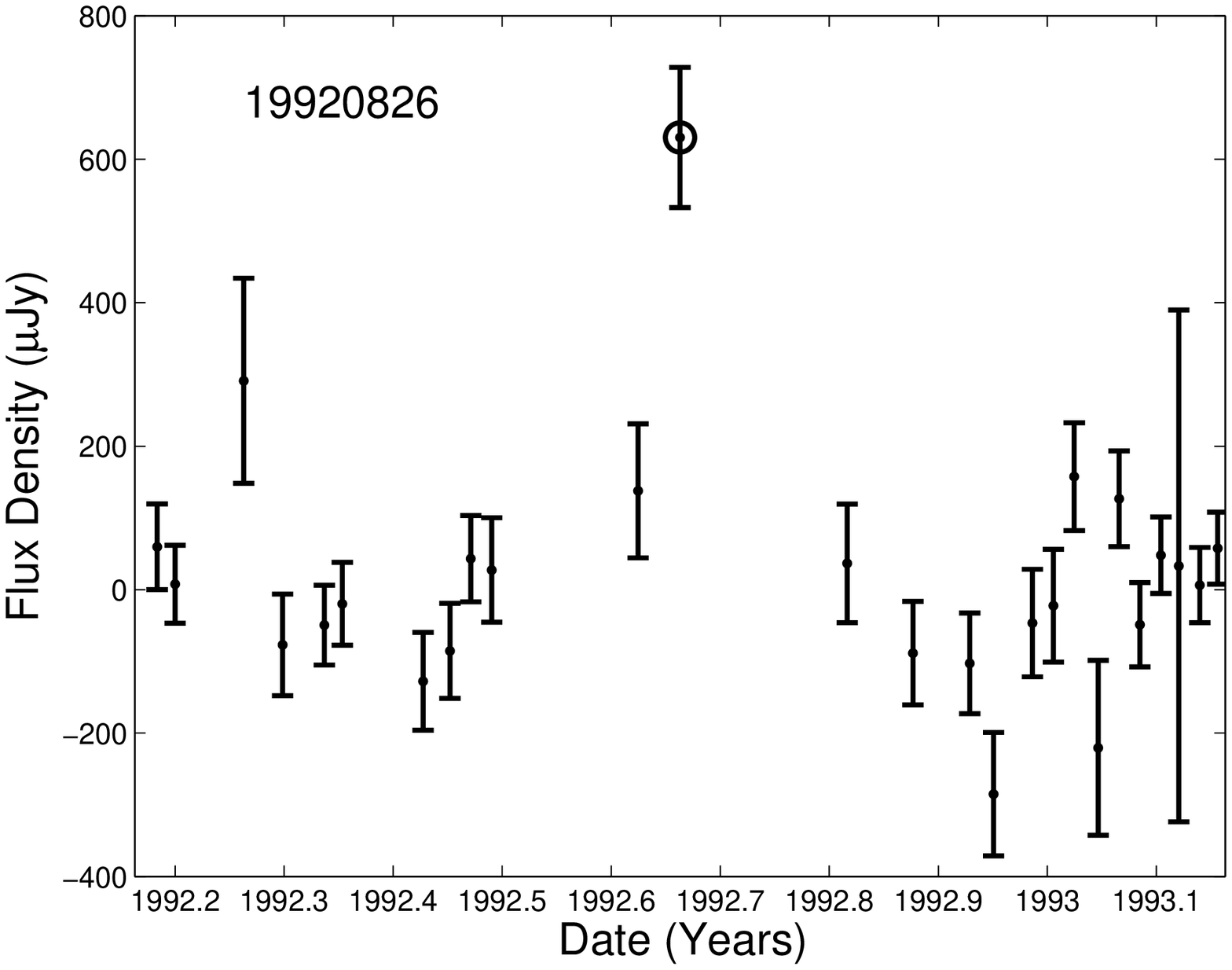}}
\mbox{\includegraphics[width=3in]{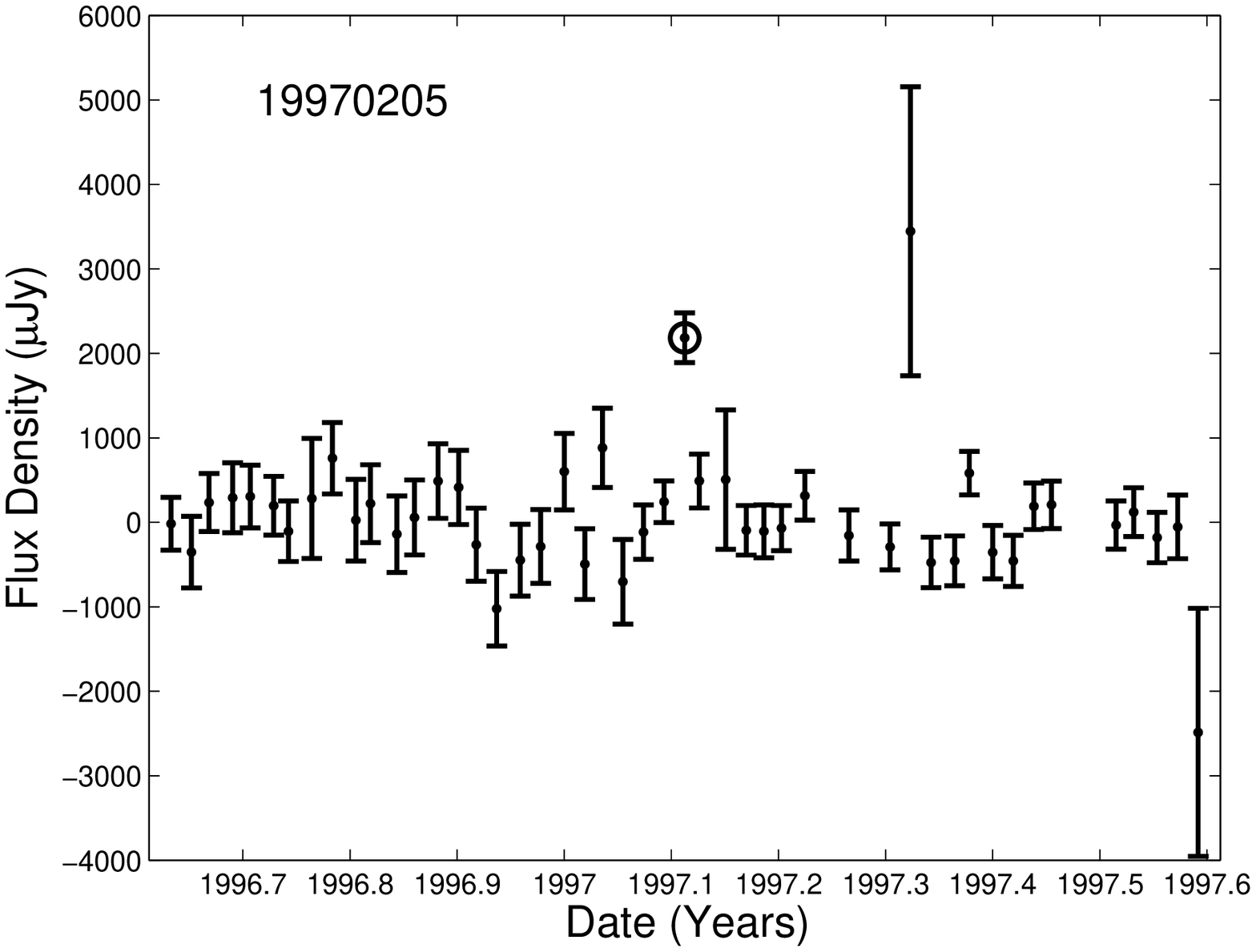}\includegraphics[width=3in]{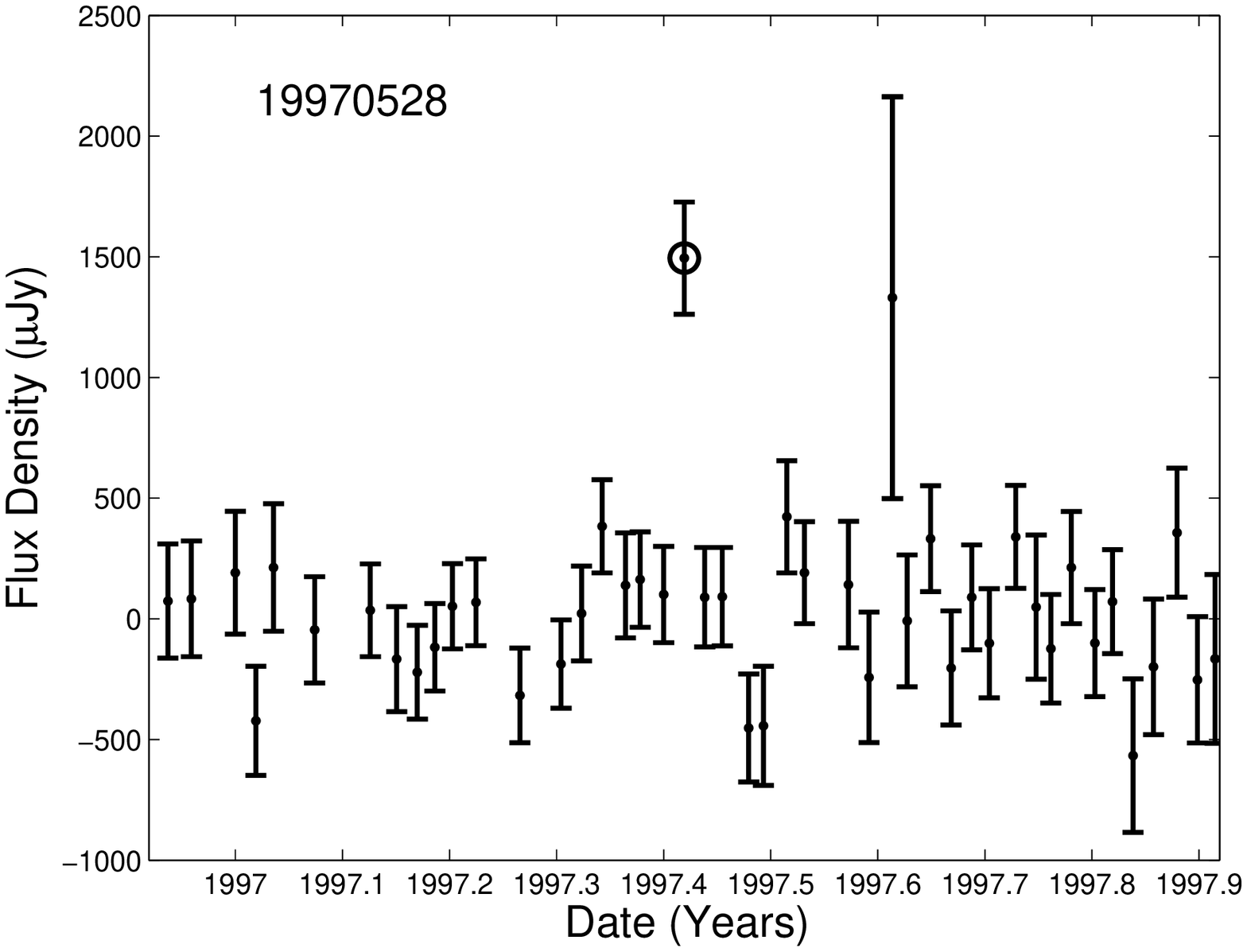}}
\mbox{\includegraphics[width=3in]{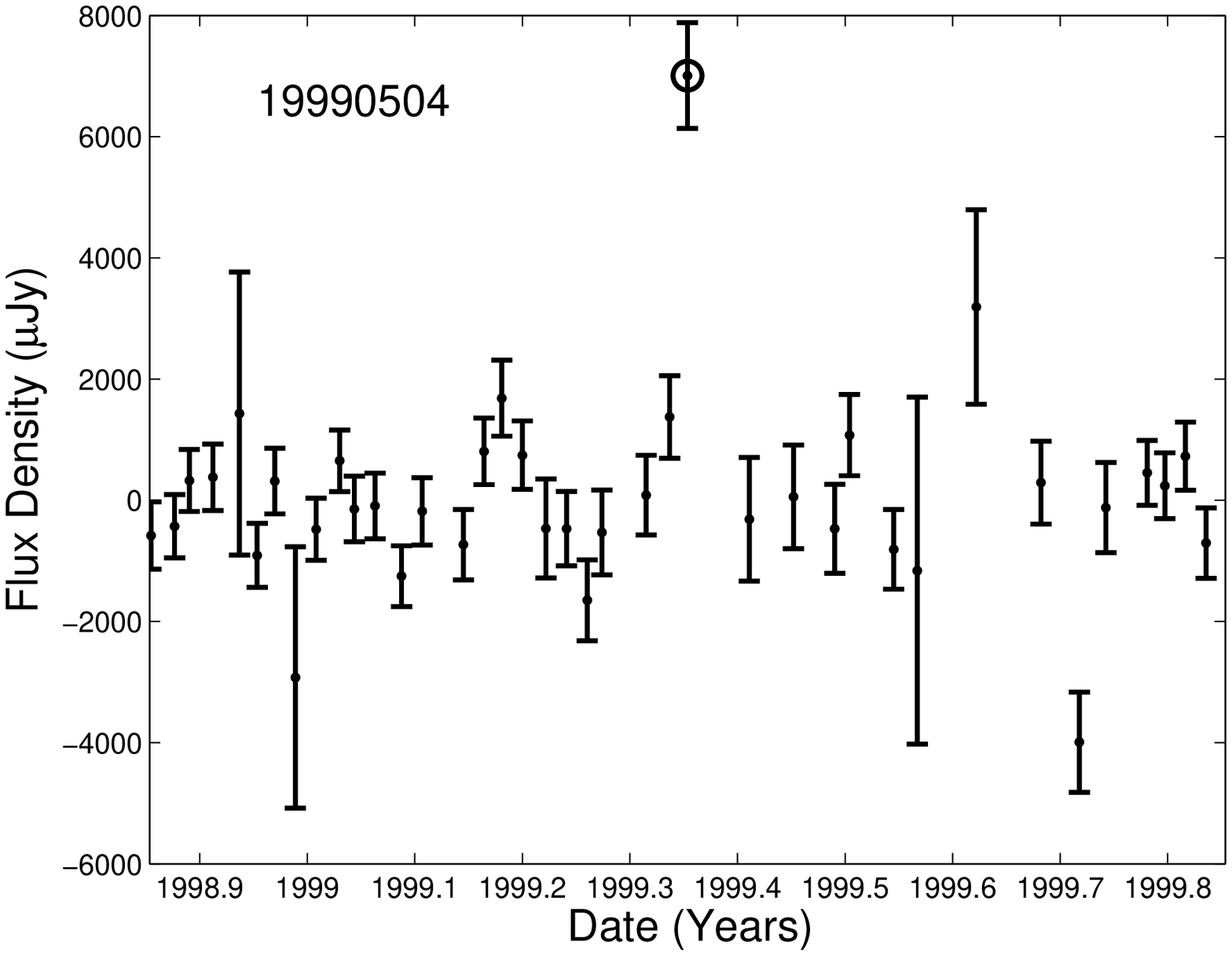}\includegraphics[width=3in]{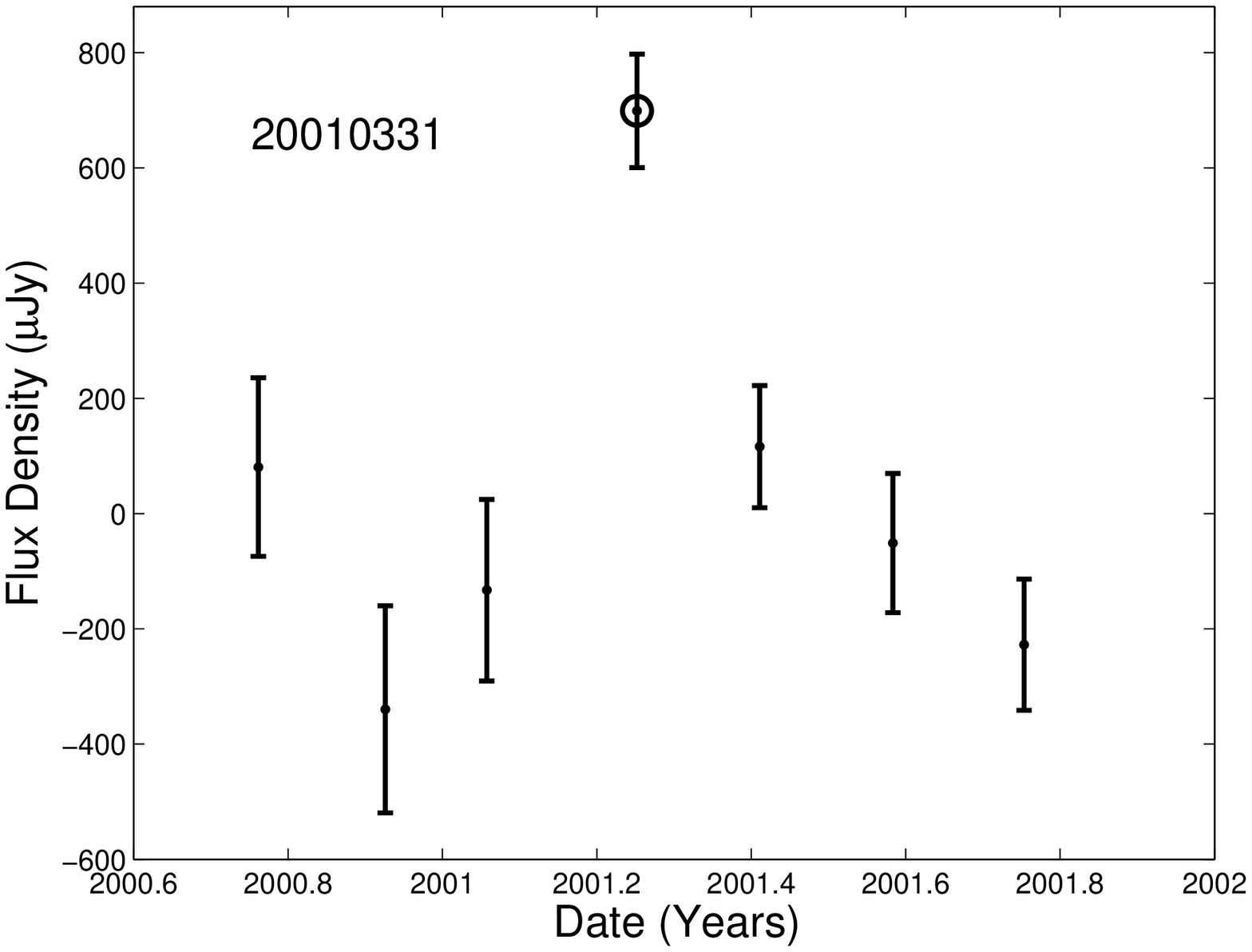}}
\figcaption{Light curves of the transient sources for the year surrounding 
detection.  A circle around the point denotes the epoch  of detection.  Other
points for all sources are all non-detections; values given are best-fit 
Gaussians at the position of the transient.
\label{fig:lct}}

\parbox[b]{\textwidth}{\includegraphics[width=2in]{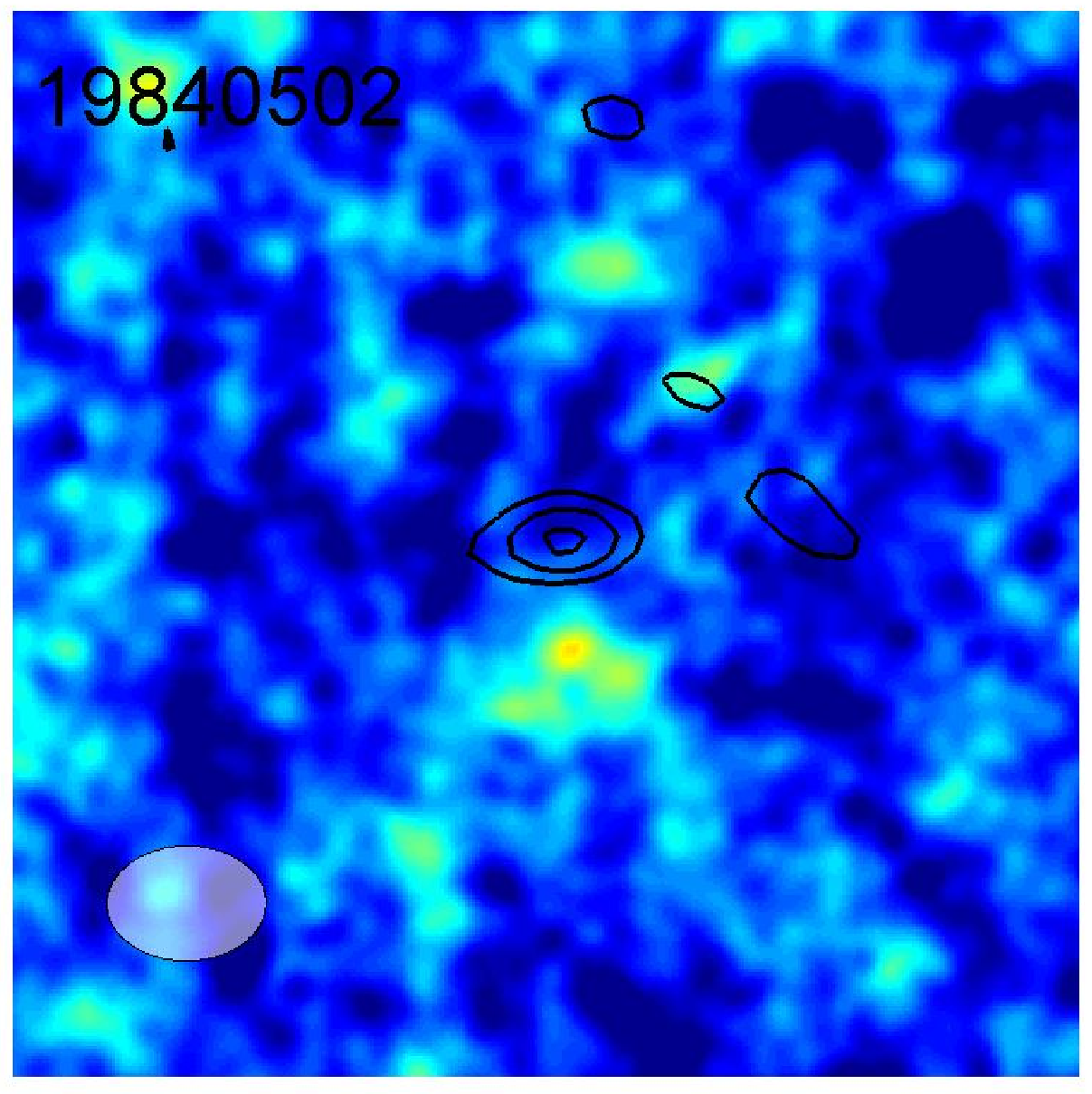}\includegraphics[width=2in]{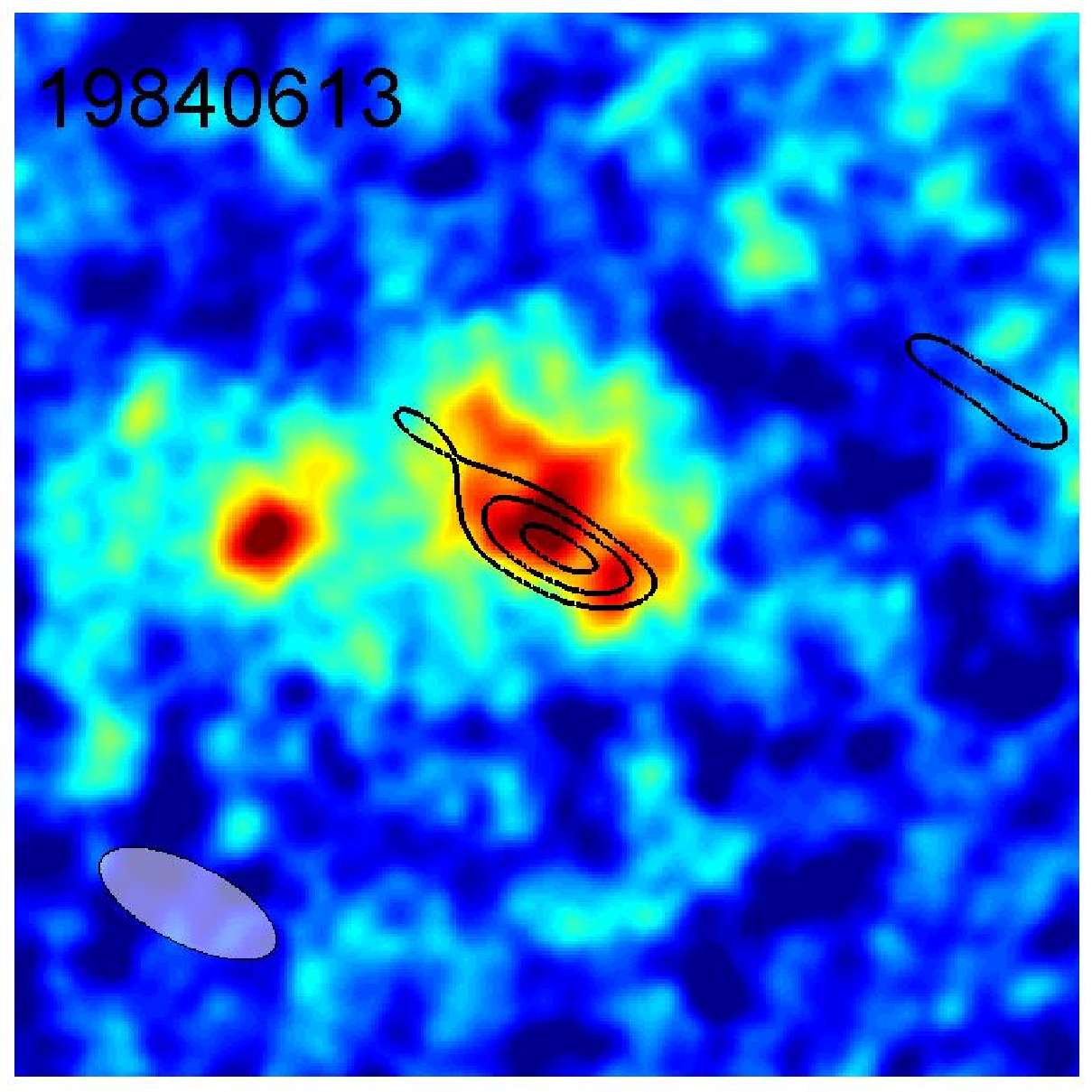}\includegraphics[width=2in]{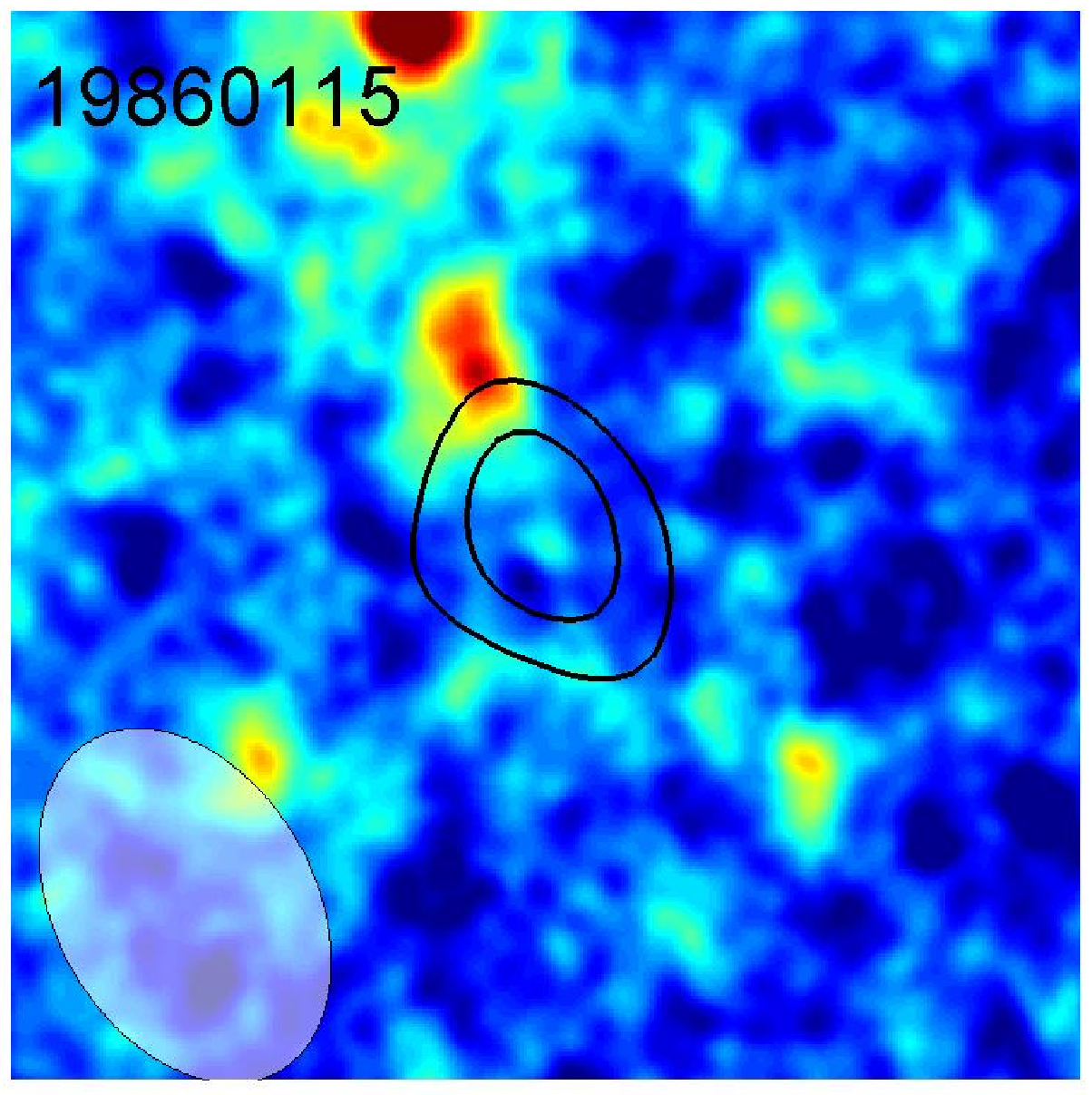} \\
\includegraphics[width=2in]{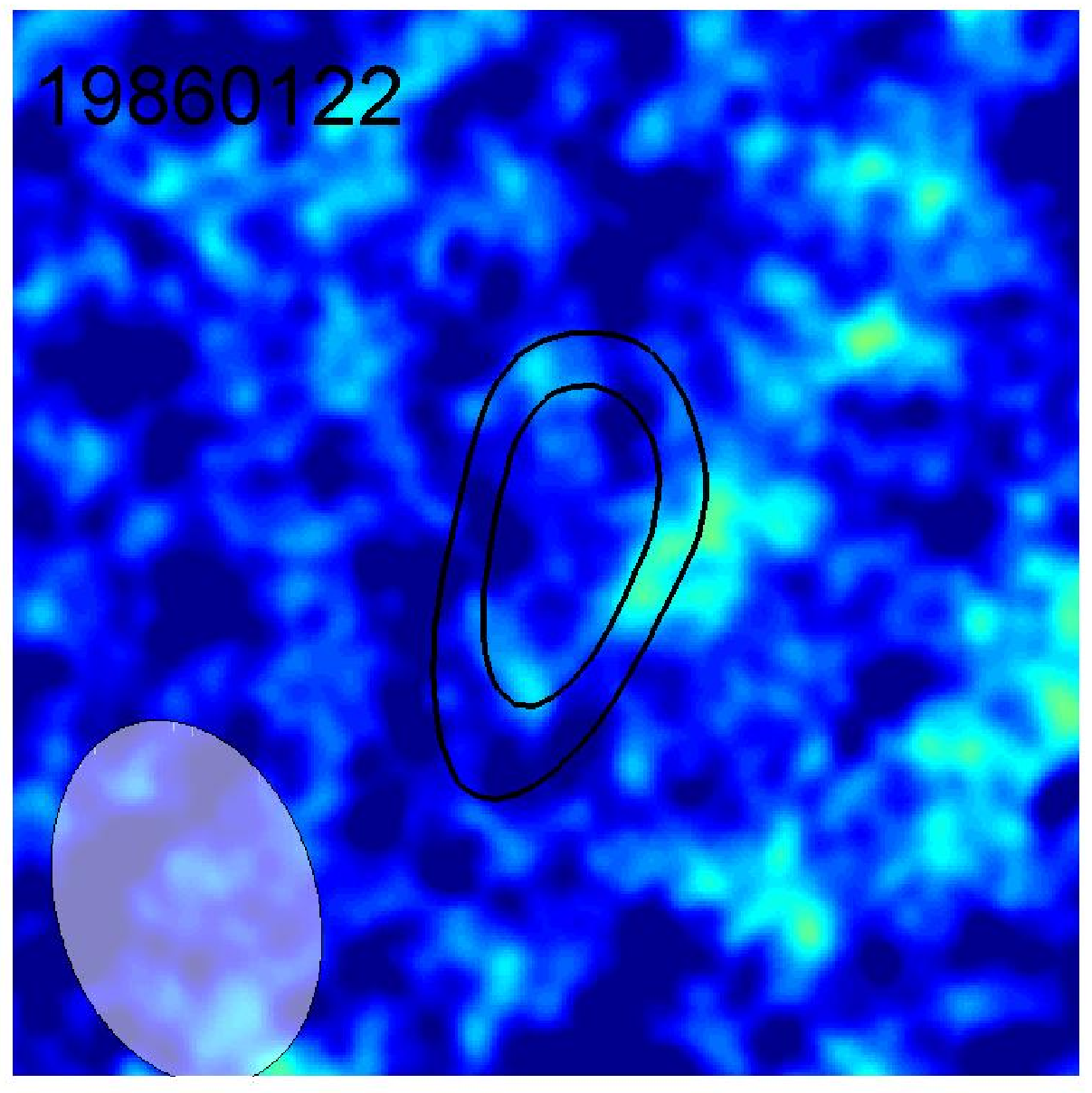}\includegraphics[width=2in]{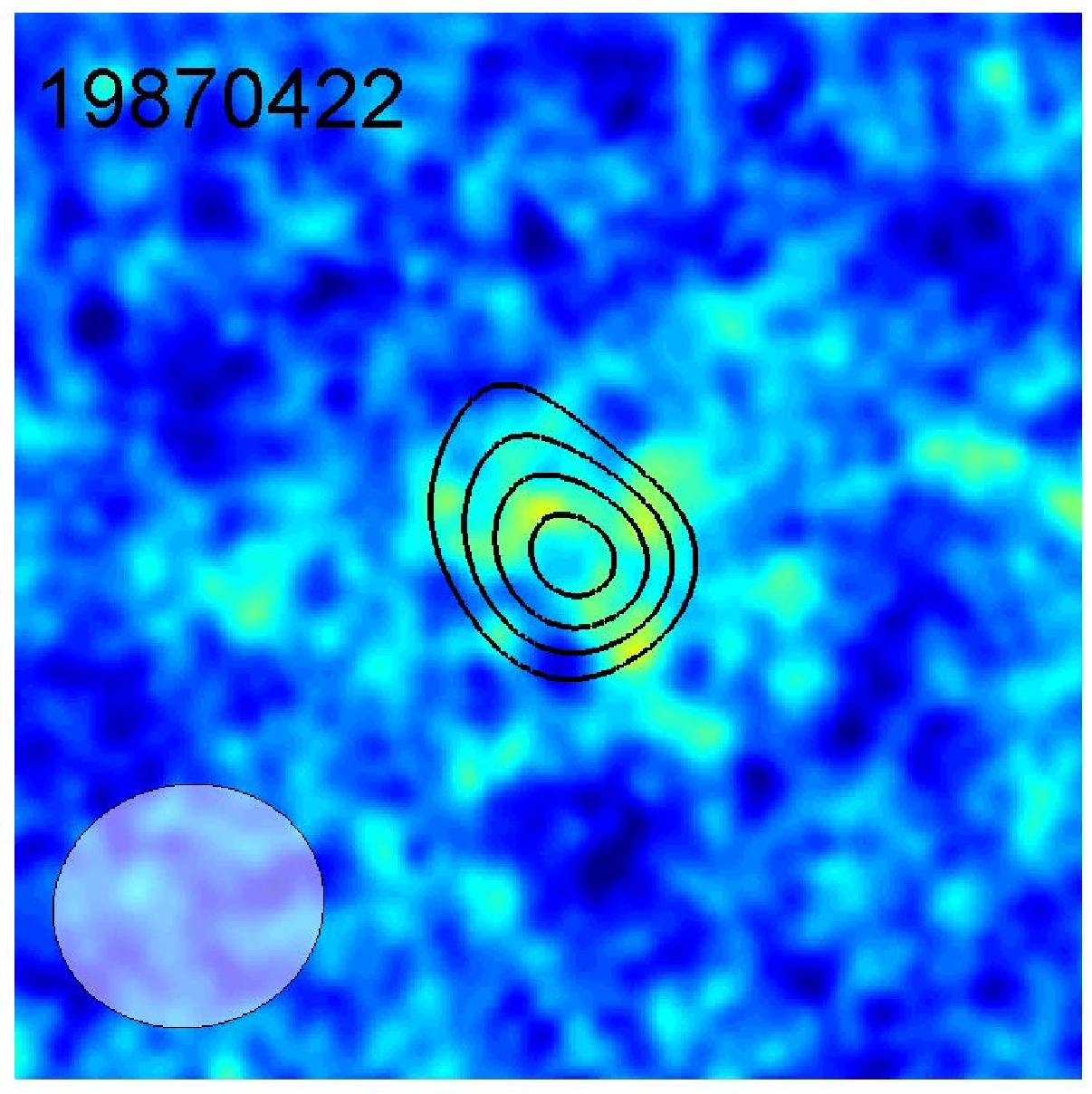}\includegraphics[width=2in]{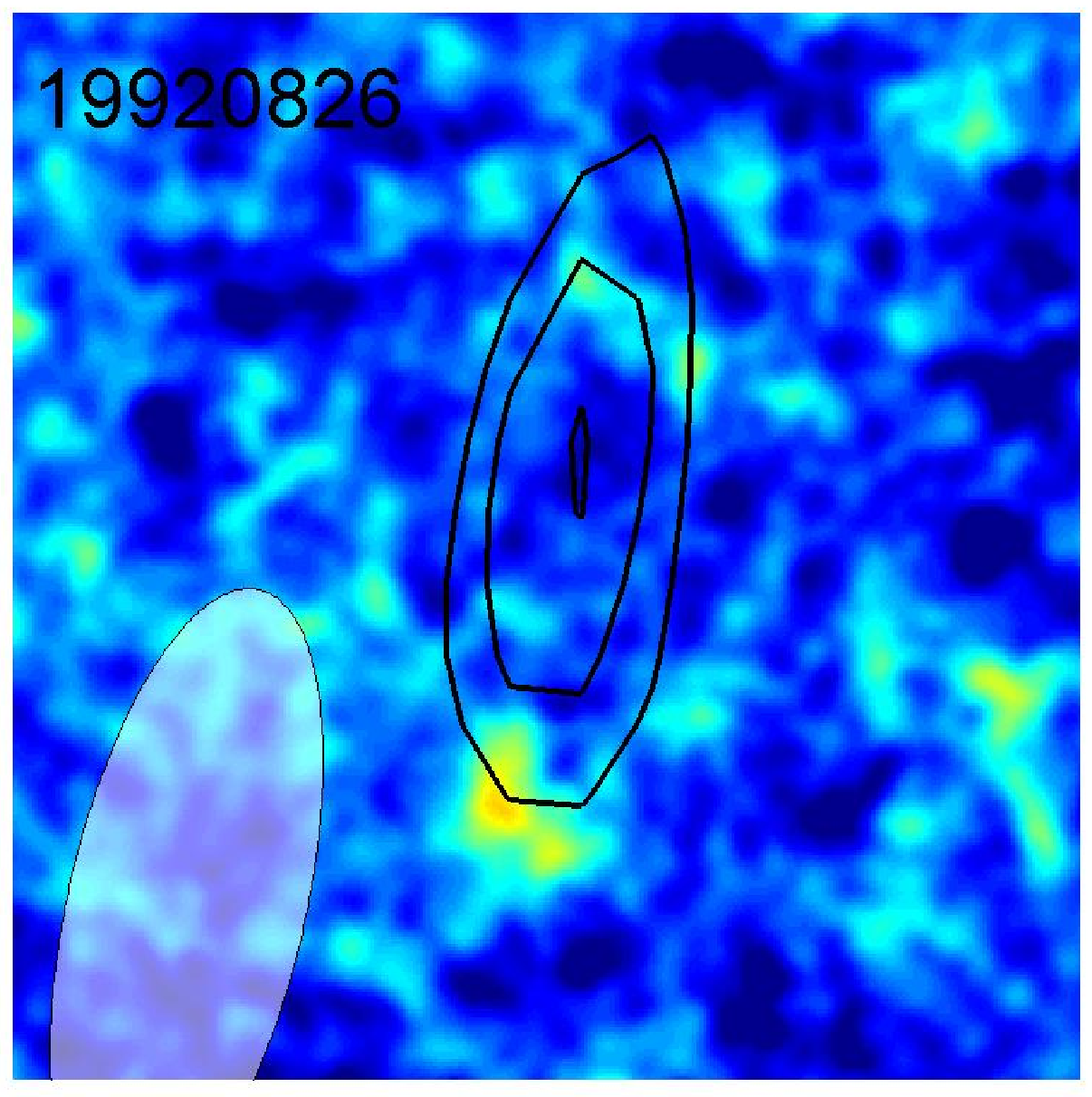} \\
\includegraphics[width=2in]{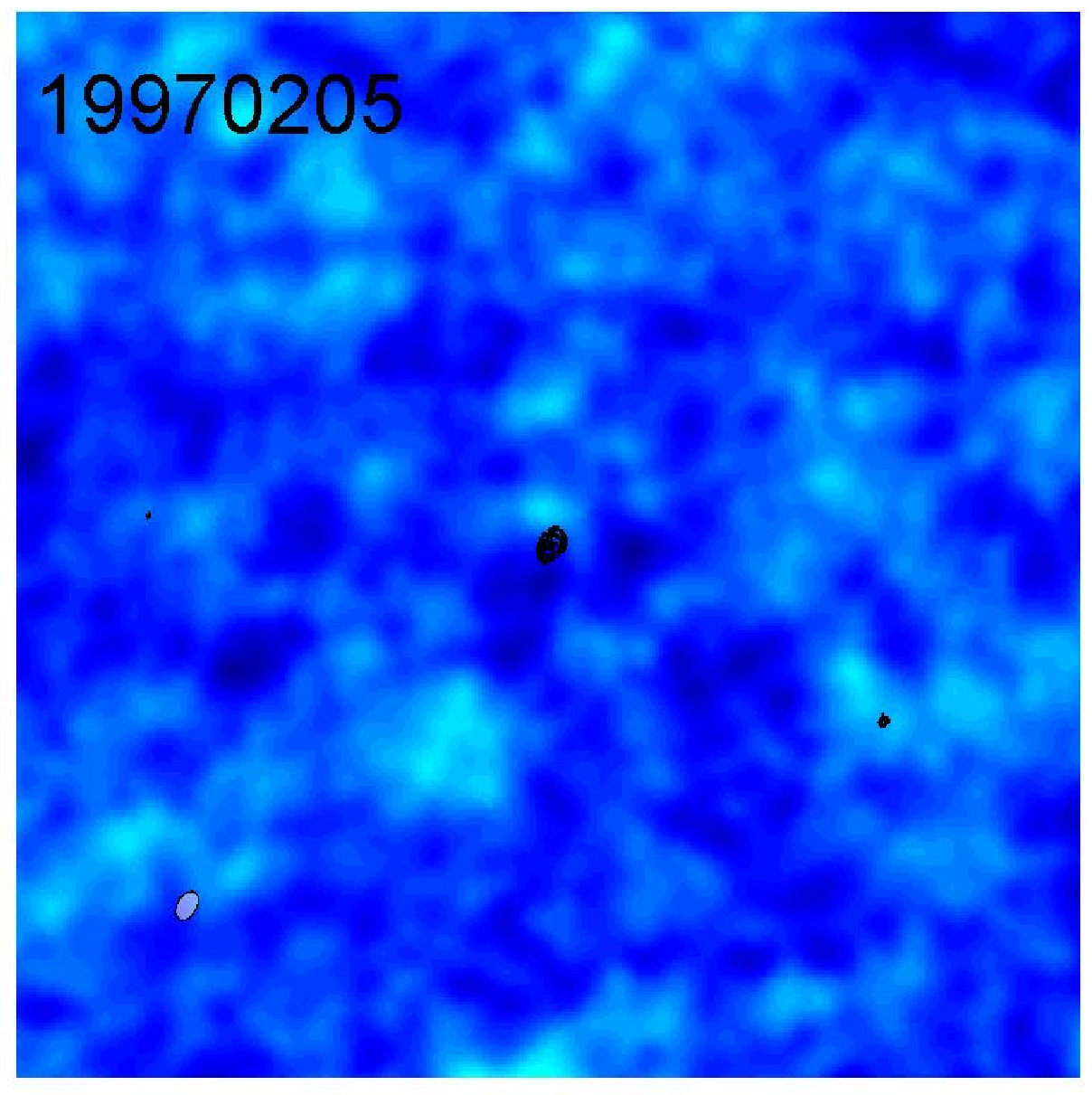}\includegraphics[width=2in]{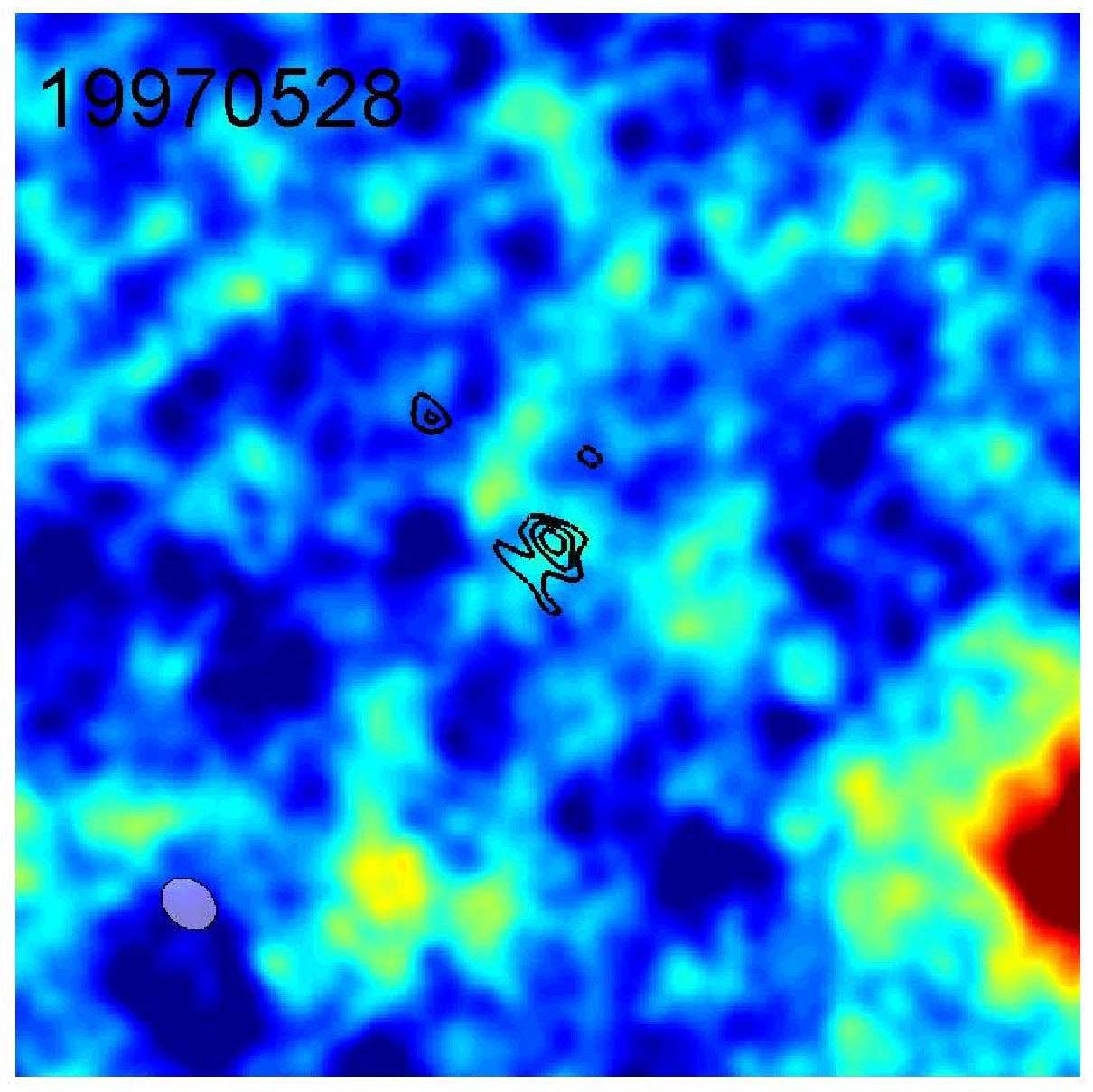}\includegraphics[width=2in]{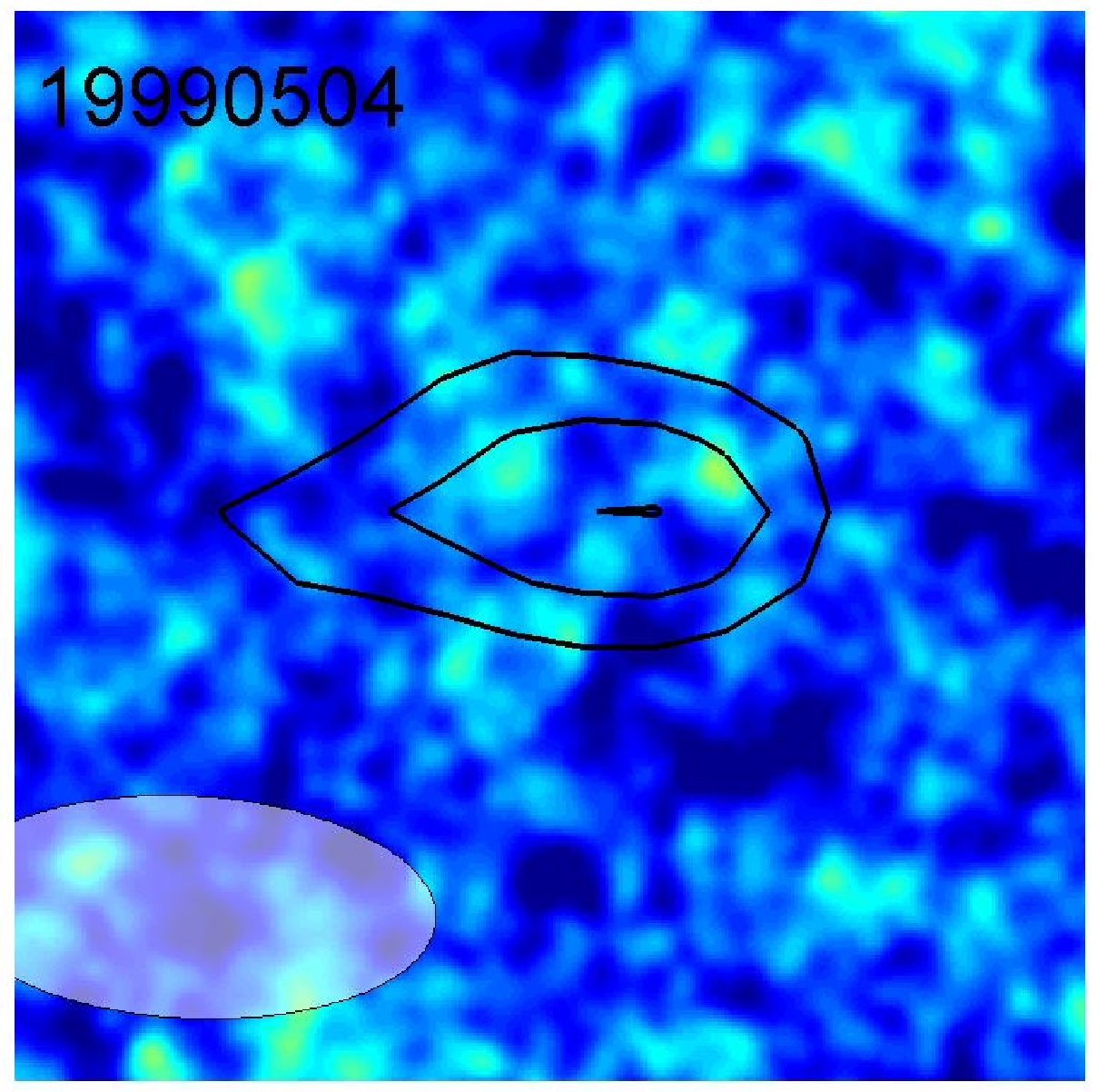} \\
\includegraphics[width=2in]{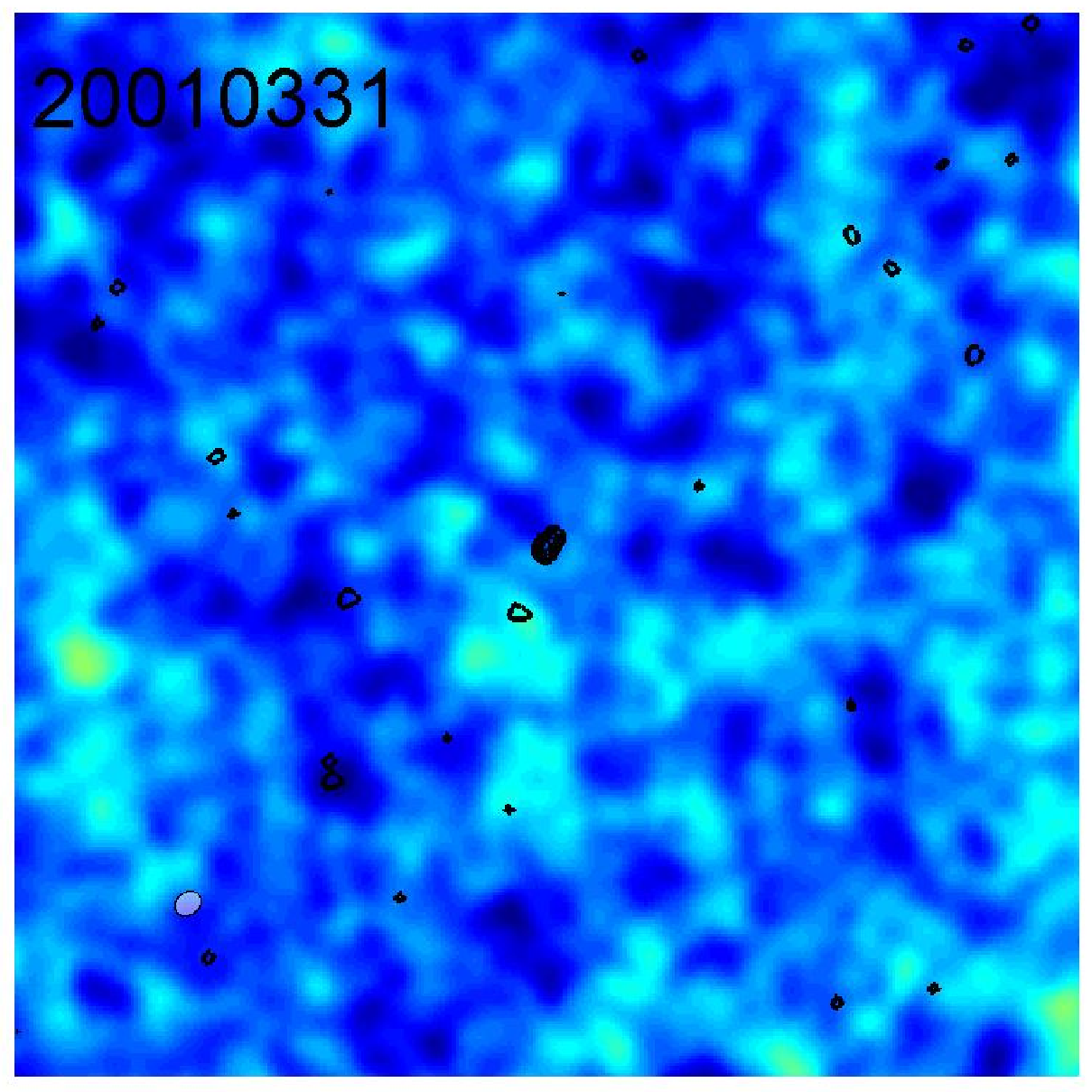}}
\figcaption{Contour plots of transients overlaid on color images from 
the deep radio image. Each image is $1'$ in scale. Contours are 3, 4, 5, 6, 
and 7 times the image rms.  The synthesized beam for each image is 
displayed in blue in the lower left-hand corner.  The color
scale of the deep radio image has a range of 0 to 100 $\mu$Jy.
\label{fig:deepradio}}

\parbox[b]{\textwidth}{\includegraphics[width=2in]{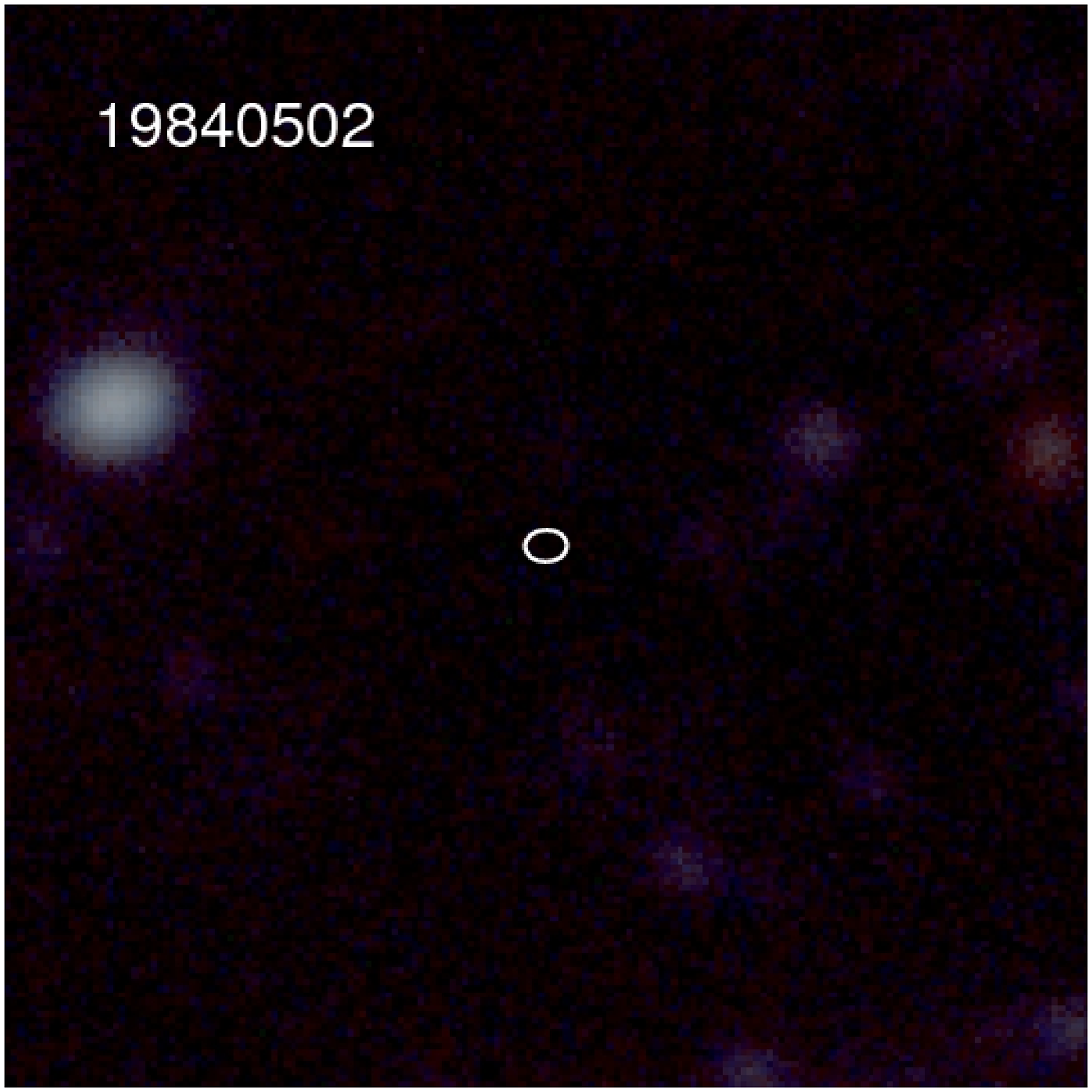}\includegraphics[width=2in]{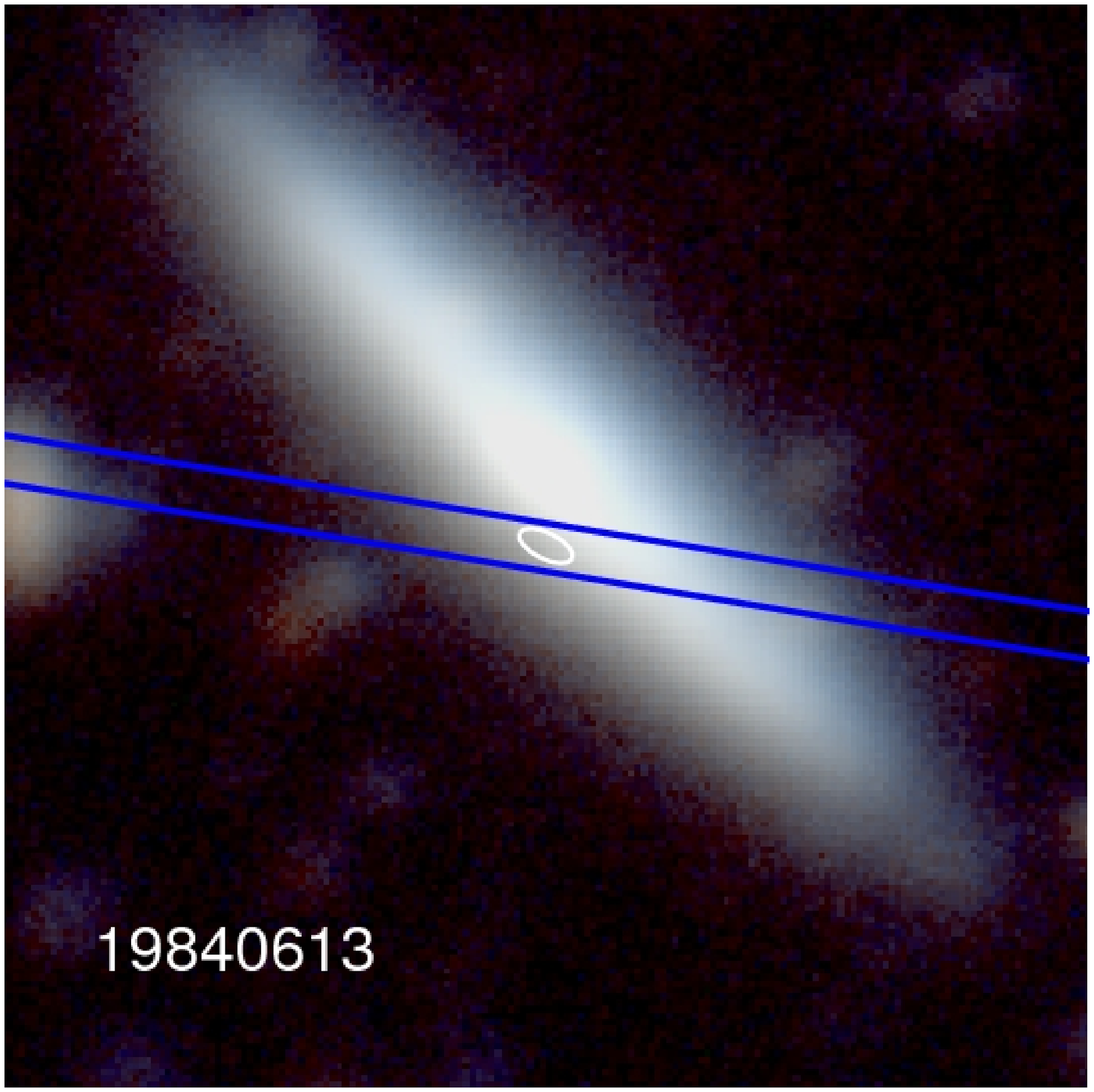}\includegraphics[width=2in]{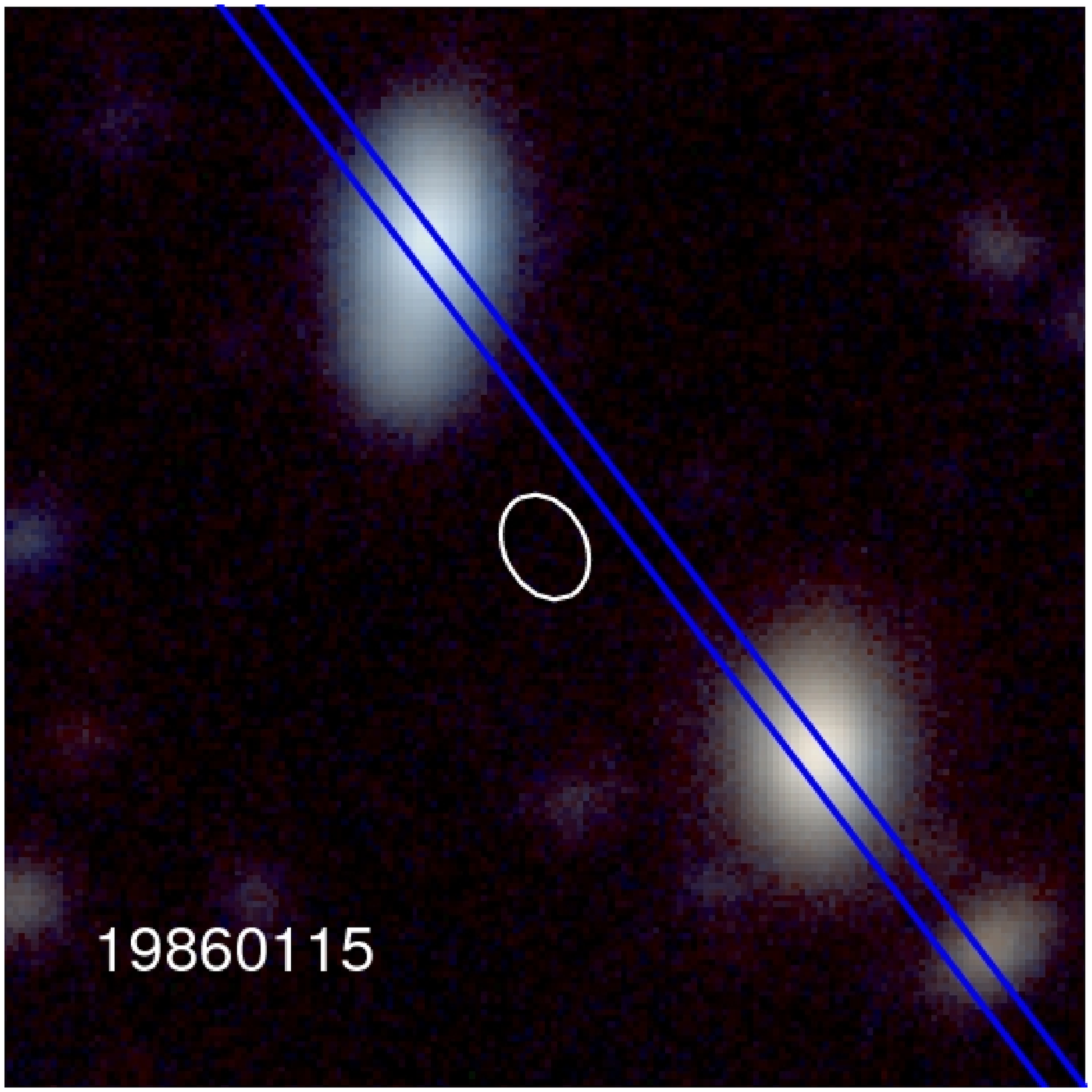} \\
\includegraphics[width=2in]{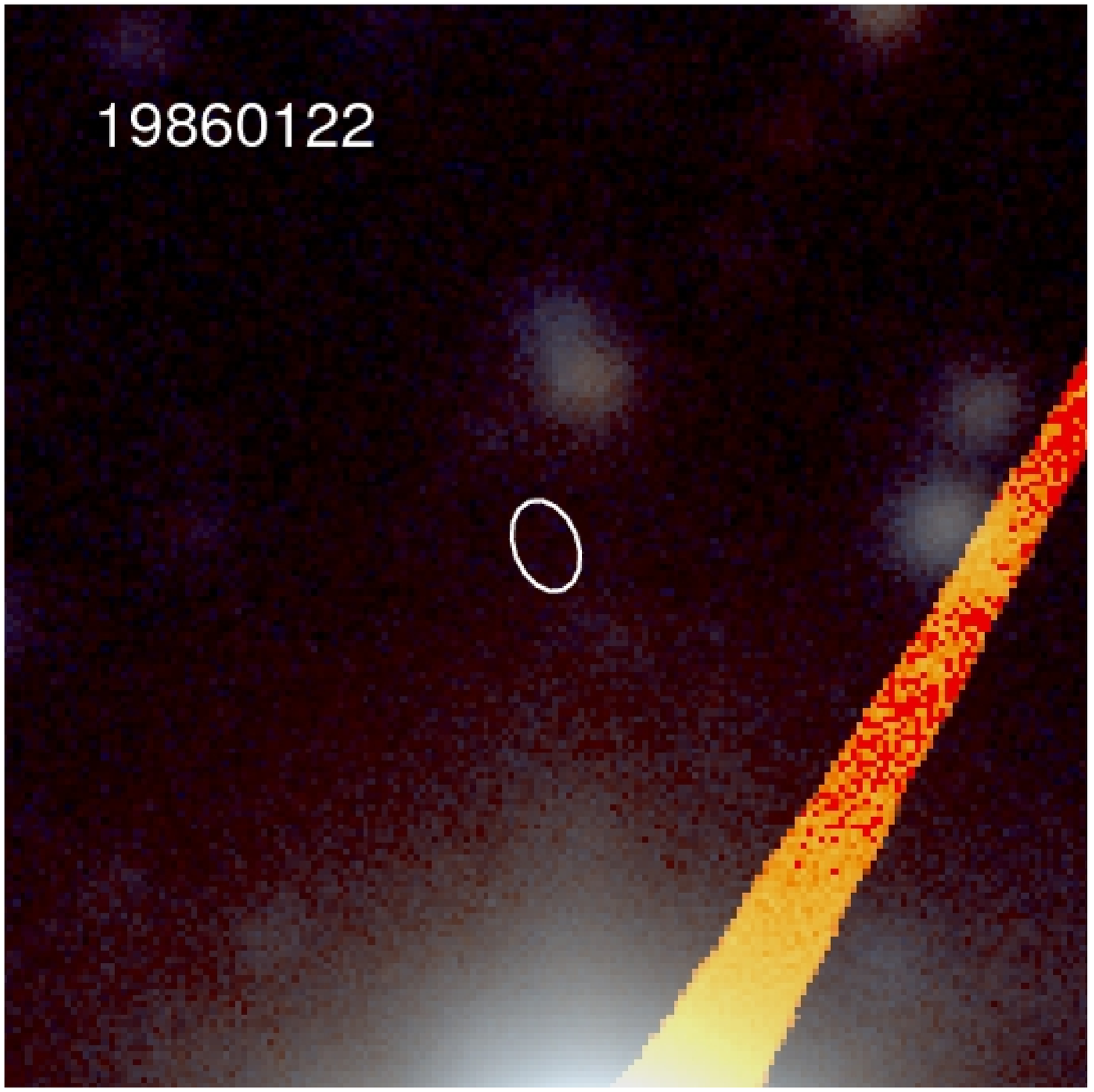}\includegraphics[width=2in]{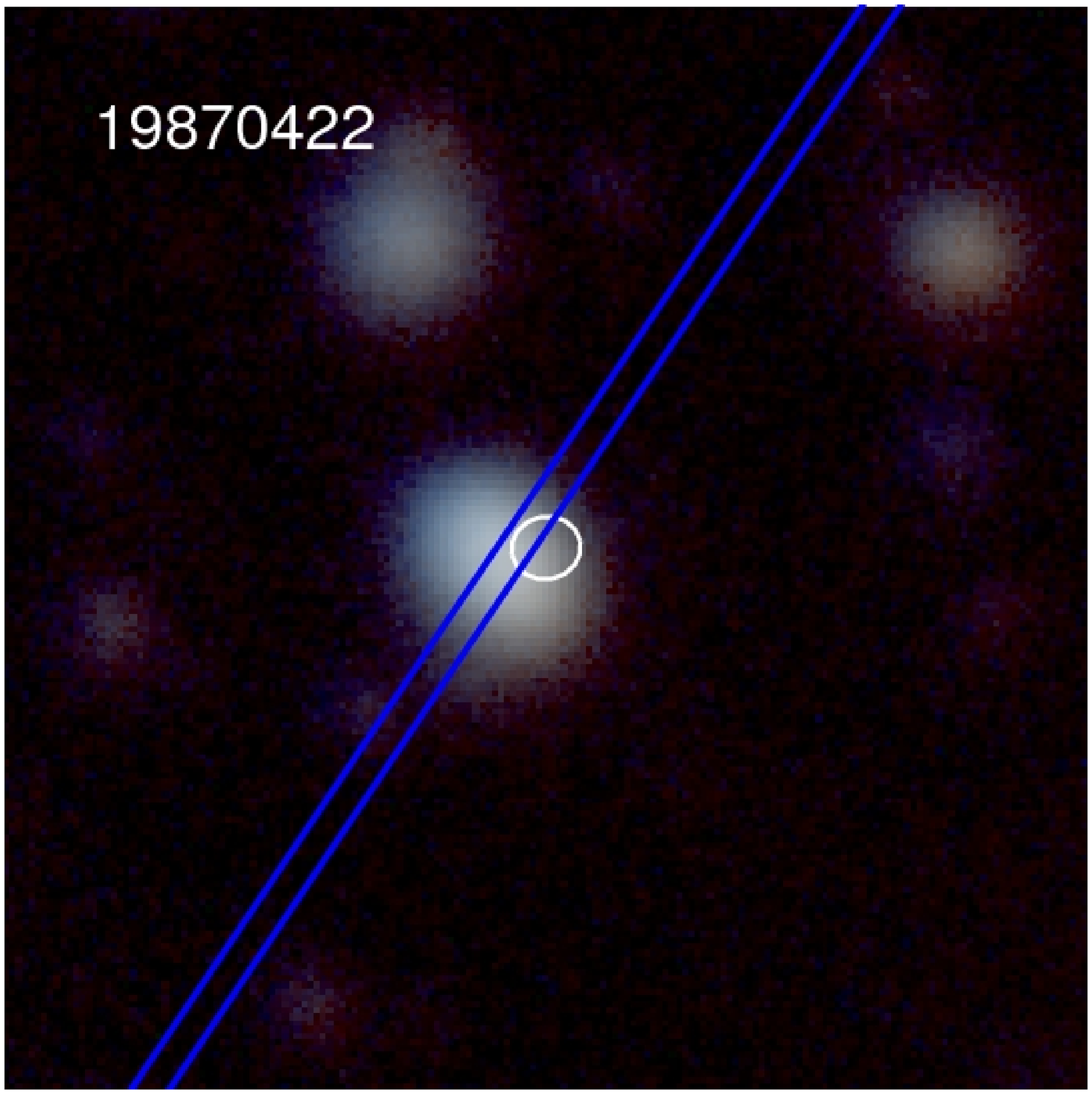}\includegraphics[width=2in]{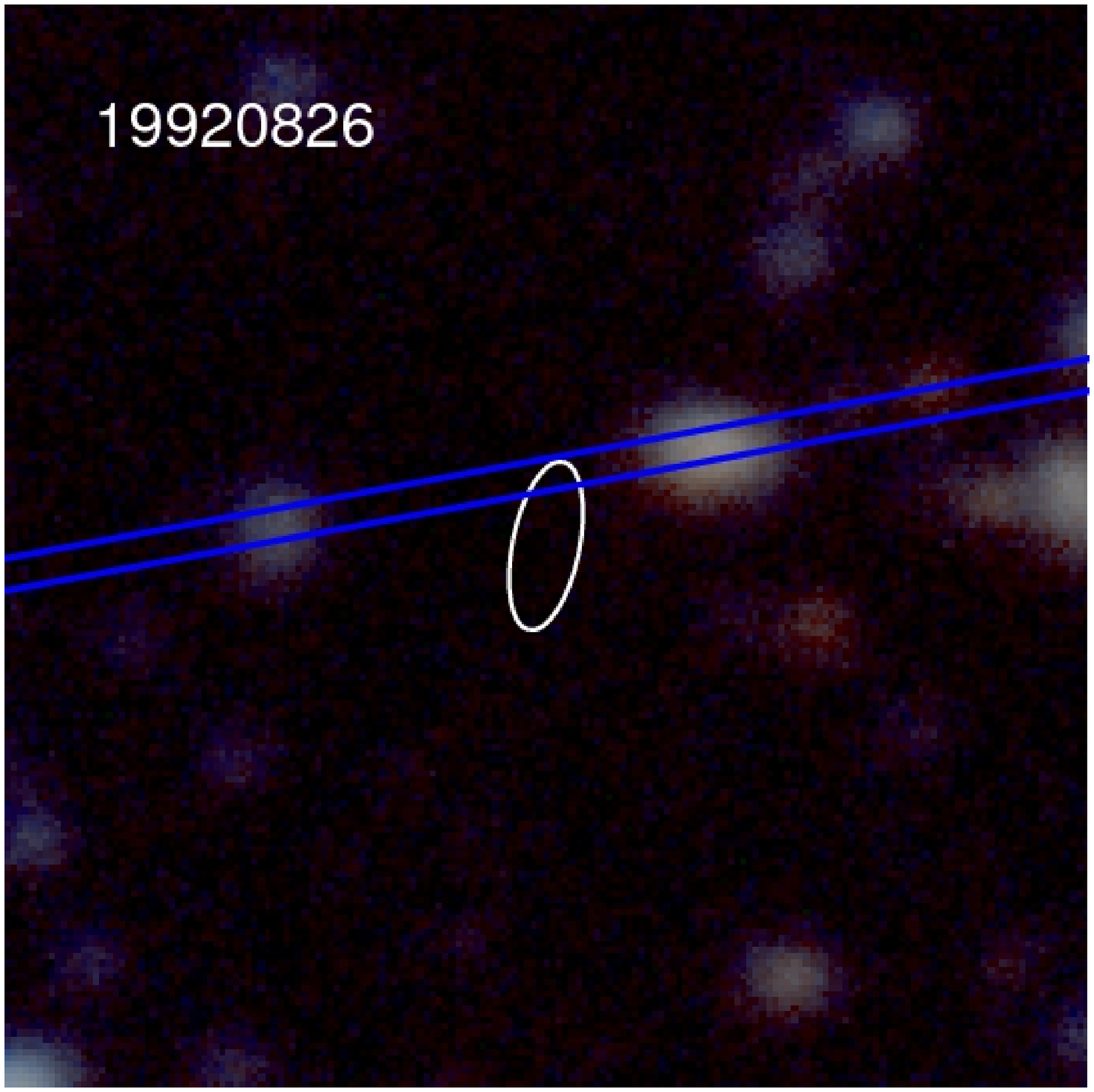} \\
\includegraphics[width=2in]{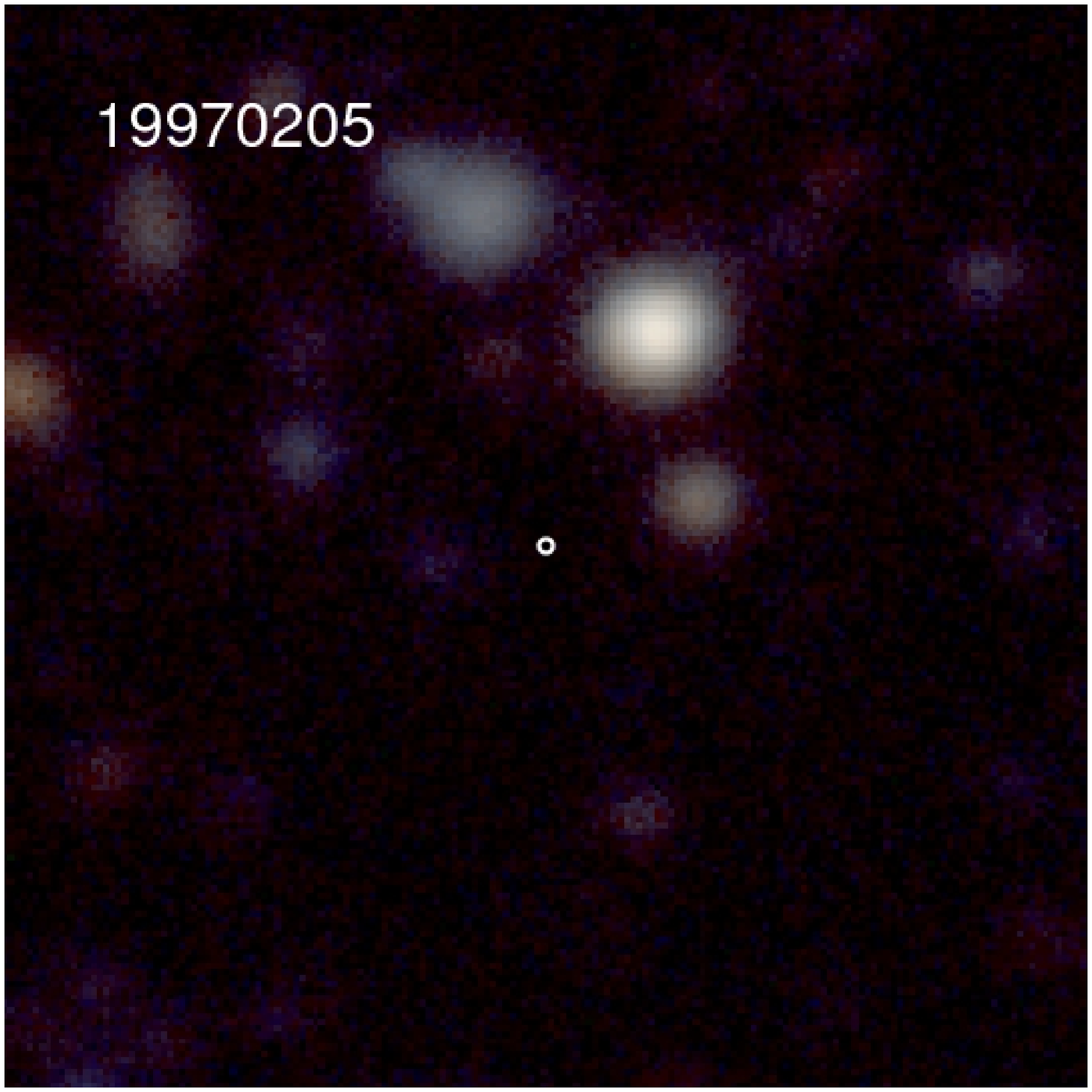}\includegraphics[width=2in]{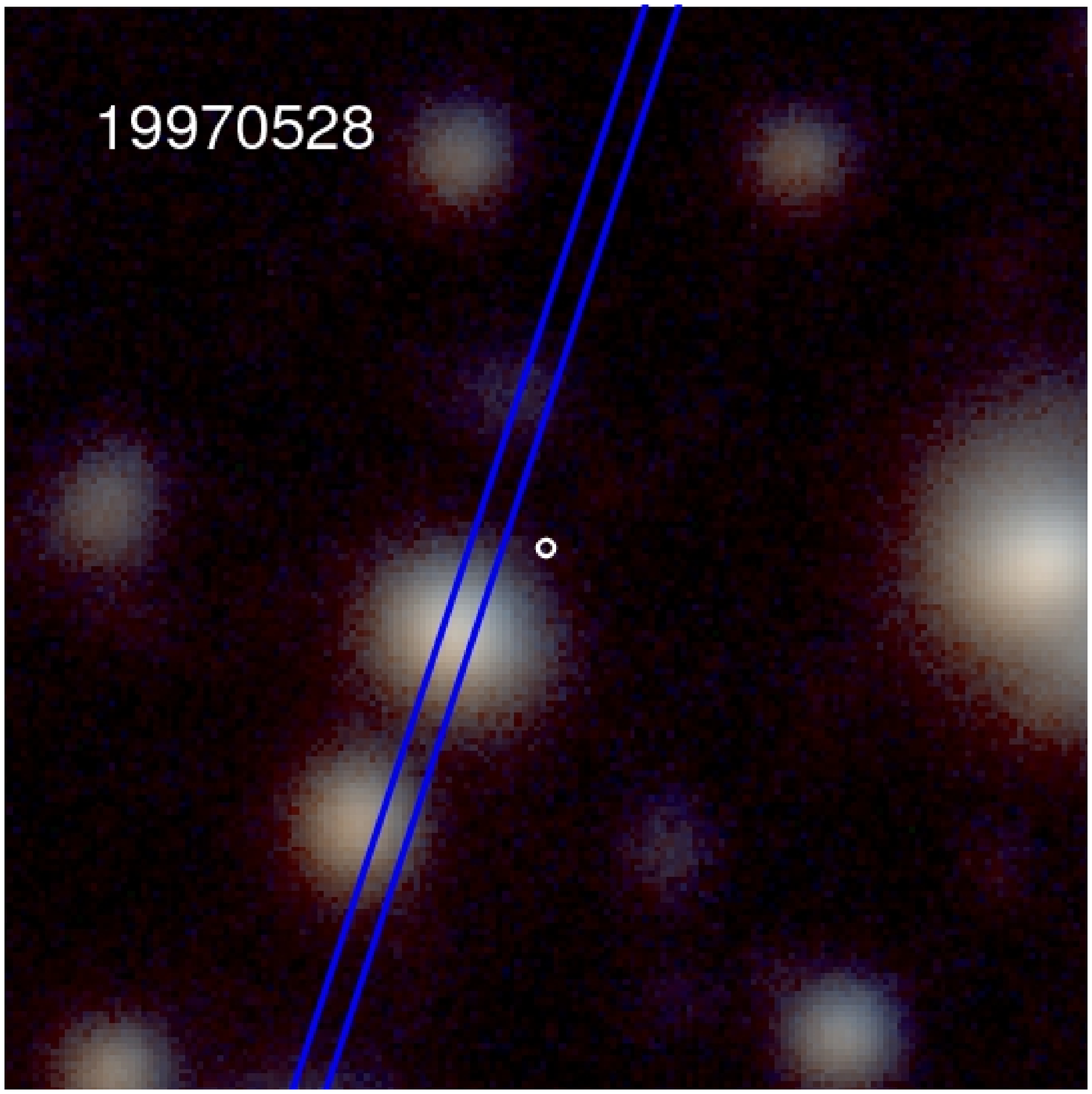}\includegraphics[width=2in]{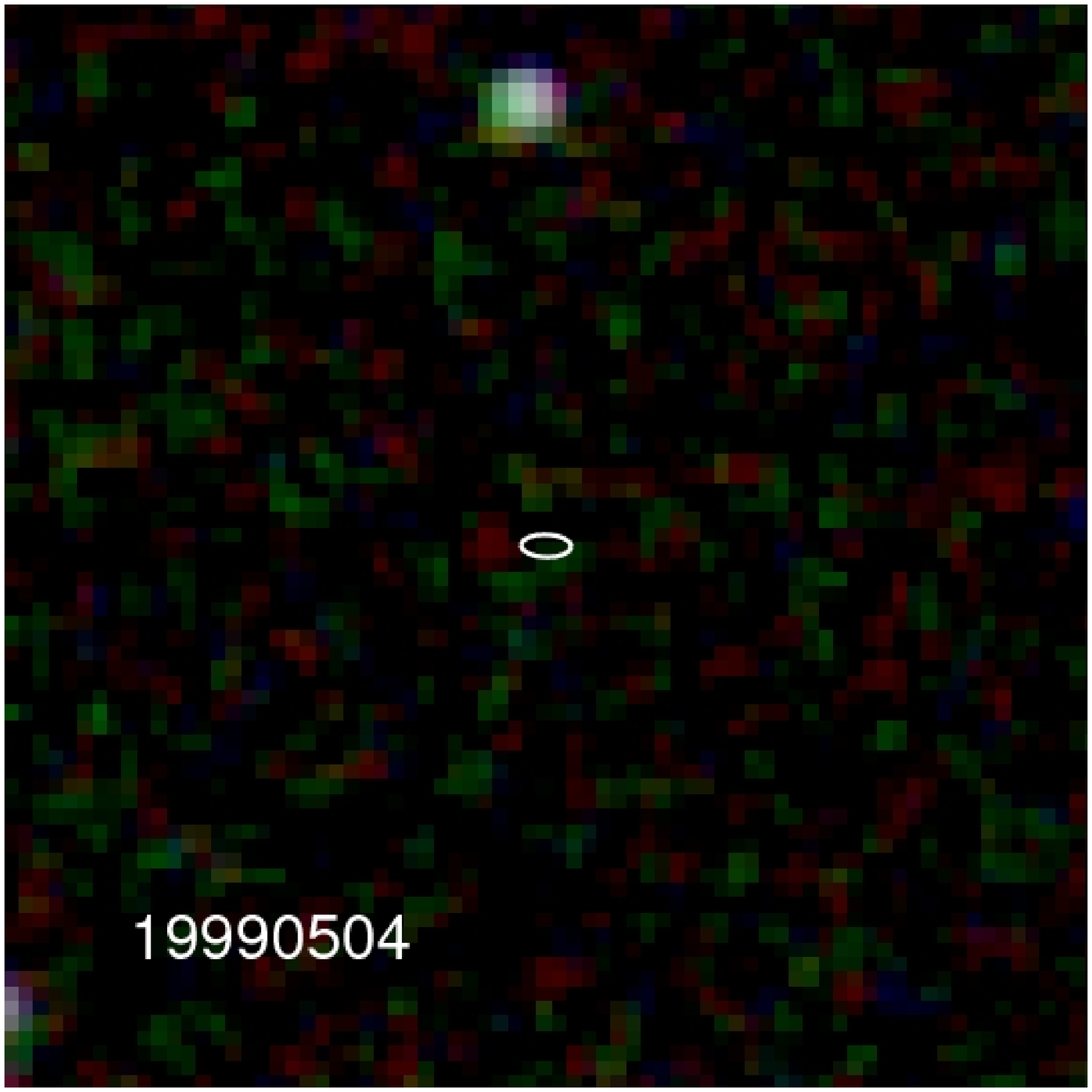} \\
\includegraphics[width=2in]{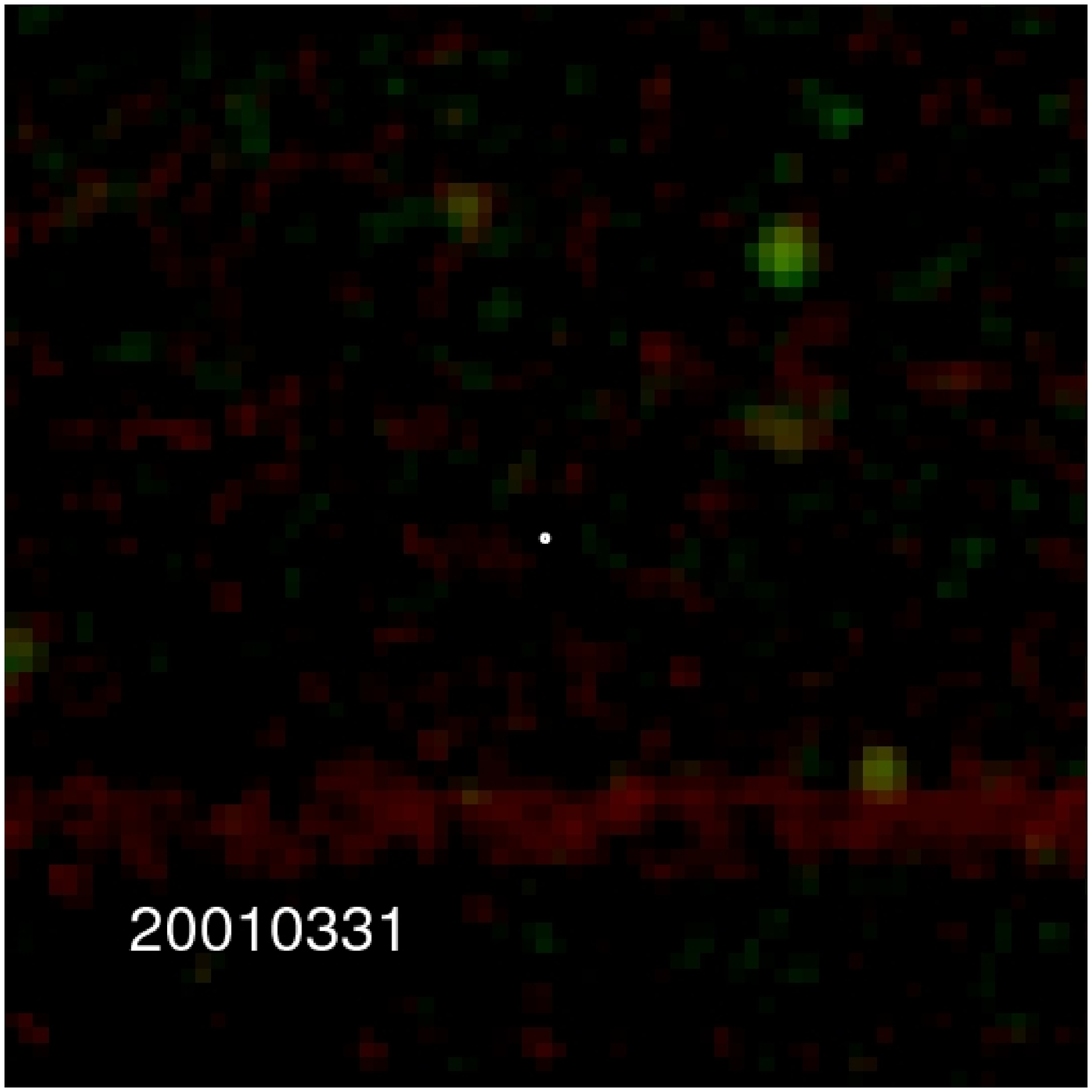}}
\figcaption{Optical and infrared images at the positions of transient 
sources.  The white circle indicates the uncertainty in the position 
of the RT.  Images are from Keck except for RTs 19990504 and 20010331,
which are from PAIRITEL.  The color composite for the Keck images
was made by using the $g$-band image for the blue channel, $R$ band 
for the red channel, and the geometric mean of the two images for the 
green channel.  Keck images are $38''$ on a side. The PAIRITEL image 
is in the $K$ band and is $73''$ on a side.  The orientation and 
location of the spectroscopic slit is indicated with blue bars.
North is up and east is to the left.  The bright streak in the image of 
RT 19860122 is a diffraction spike from a nearby bright star.
\label{fig:keck}}

\includegraphics[angle=-90,width=6in]{f7.ps}
\figcaption{Deep image of the region centered on the transient 
19840613.  The small cross marks the position and $3\sigma$ uncertainty 
in its position.  The position is clearly offset from the centroid of 
the radio emission and from the centroid of the optical emission, marked 
with the large cross.  The extent of the large cross delineates the 
major and minor axes of the galaxy as measured in 2MASS. Contours are 
3, 4, 5, 6, 7, 8, 9, and 10 times the image rms of 2.6 $\mu$Jy.
\label{fig:840613deep}}

\includegraphics[angle=-90,width=6in]{f8.ps}
\figcaption{Deep image of the region centered on the transient 
19870422.  We mark the position and $3\sigma$ uncertainty of the 
RT (plus sign) and the position of the optical centroid (star).  
The RT position is offset from any peak of the radio emission 
but is consistent with the peak of the optical emission.   
Contours are $-2$, 2, 3, and 4 times the image rms of 14~$\mu$Jy 
at the location of the transient.
\label{fig:870422deep}}

\plotone{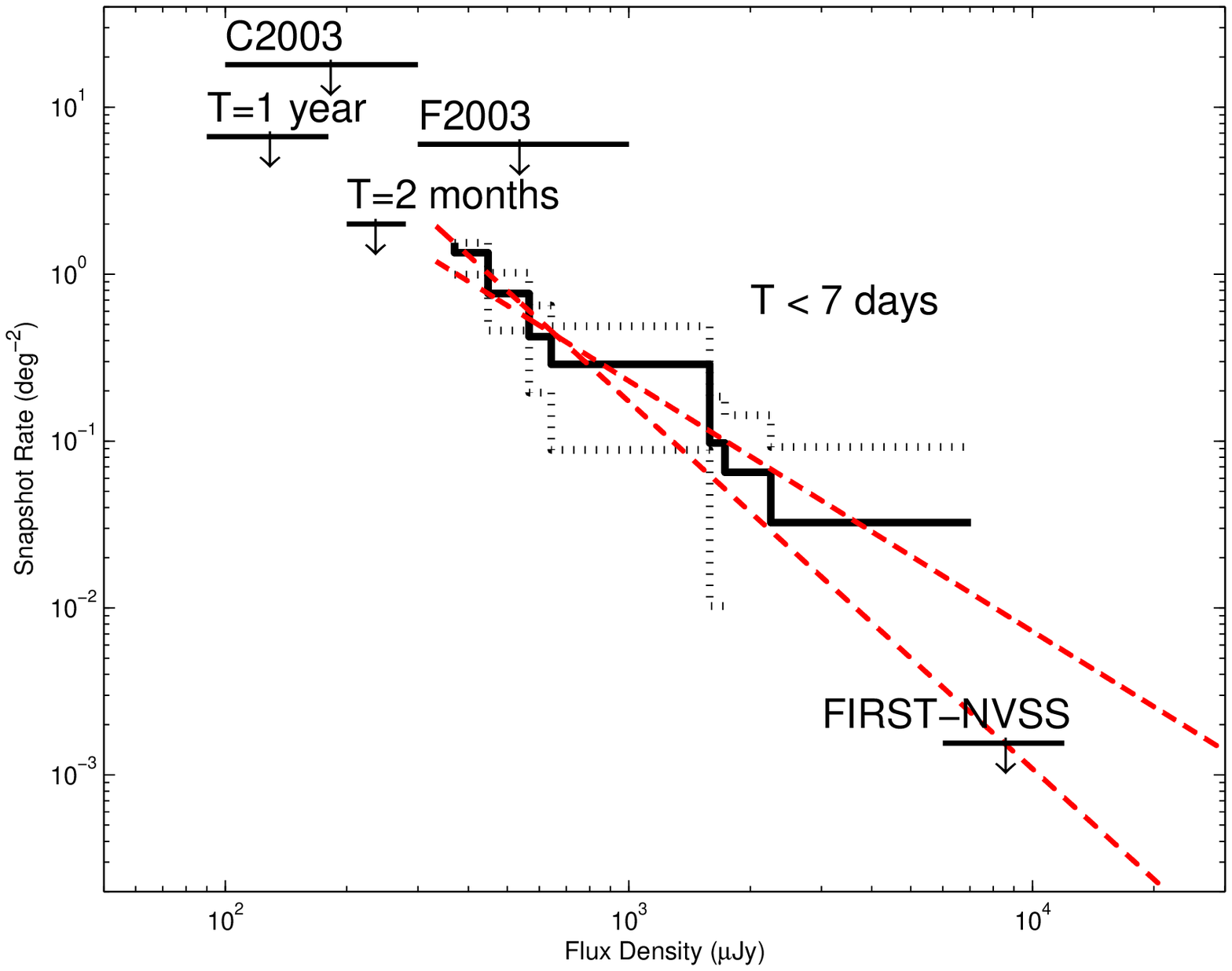}
\figcaption[]{The cumulative
two-epoch source density for radio transients as a function of 
flux density.  The solid black line shows the rate while the 
dotted lines show the $2\sigma$ upper and lower bounds. The red 
dashed lines show $S^{-1.5}$ and $S^{-2.2}$ curves.  Both are 
reasonable fits to these data, while the latter shows greater 
consistency with lower-frequency results. The arrows show 
$2\sigma$ upper limits for transients from this survey with a 
1-year timescale, two-month timescale, and for transients 
from the comparison of the 1.4~GHz NVSS and FIRST surveys 
\citep{2006ApJ...639..331G}, from the \citet{2003ApJ...590..192C} 
survey (labeled C2003), and from the \citet{2003AJ....125.2299F} 
survey (labeled F2003).  See the main text for details of these 
surveys.
\label{fig:snapshotrate}}

\end{document}